\documentclass[10pt]{article}

\usepackage{amsmath}
\usepackage{amssymb}

\usepackage{graphicx}

\usepackage{cite}

\usepackage{color} 
\usepackage{subcaption}

\usepackage[labelfont=bf,labelsep=period,justification=raggedright]{caption}

\makeatletter
\renewcommand{\@biblabel}[1]{\quad#1.}
\makeatother

\date{}

\begin{document}

% Title must be 150 characters or less
\begin{flushleft}
{\Large
%\textbf{Hierarchy of financial markets data: comparison of clustering methods}
\textbf{\color{black} Relation between financial market structure and the real economy:  comparison between clustering methods %and industrial sectors
}
}
\vskip.2cm
% Insert Author names, affiliations and corresponding author email.
Nicol\'o Musmeci$^{1}$, 
Tomaso Aste$^{2,3,\ast}$, 
T. Di Matteo$^{1}$
\vskip.2cm
\bf{1} Department of Mathematics, King's College London, The Strand, London, WC2R 2LS, UK
\\
\bf{2} Department of Computer Science, UCL, Gower Street, London, WC1E 6BT, UK
\\
\bf{3} Systemic Risk Centre, London School of Economics and Political Sciences, London, WC2A2AE, UK
\\
$\ast$ E-mail: t.aste@ucl.ac.uk
\end{flushleft}

% Please keep the abstract between 250 and 300 words
\section*{Abstract}
We quantify the amount of information filtered  by different hierarchical clustering methods on correlations between stock returns comparing the clustering structure with the underlying industrial activity classification.
%We compare the hierarchical and clustering structure from correlations between stock returns with the underlying industrial activity structure.
We apply, for the first time to financial data, a novel hierarchical clustering approach, the Directed Bubble Hierarchical Tree and we compare it with other methods including the Linkage and k-medoids. 
%In particular, b
By taking the industrial sector classification of stocks as a benchmark partition, we evaluate how the different methods retrieve this classification.   
The results show that the Directed Bubble Hierarchical Tree can outperform  other methods, being able to retrieve more information with fewer clusters.   
 Moreover, we show that the economic information is hidden at different levels of the hierarchical structures depending on the clustering method. 
 The dynamical analysis on a rolling window also reveals that the different methods show different degrees of sensitivity to  {\color{black} events affecting financial markets}, like crises. 
 These results can be of interest for all the applications of clustering methods to portfolio optimization and risk hedging.

\section*{Introduction}
\label{sec:intro}
Correlation-based networks have been extensively used in Econophysics
as tools to filter, visualise and analyse financial market data \cite{mantegna1,asset_graphs,PMFG2,Tumminello05,DiMatteo02,DiMatteo04,DiMatteo05,bartolozzi2007multi}.
Since the seminal work of Mantegna on the Minimum Spanning Tree (MST) \cite{mantegna1} they have provided insights into several aspects of financial markets including financial crises \cite{black_monday,cluster_portfolio,mst_exch_rate,pozzi_dyn_net,time_horizons,raffo_model,NJP10}. 
 The MST is strictly related \cite{mantegna2} to a hierarchical clustering algorithm, namely the Single Linkage (SL) \cite{cluster_book}. Starting from a set of elements (e.g., stocks)
 and a related distance matrix (e.g., a convenient transformation of the stocks correlation matrix 
\cite{mantegna1}), the SL performs an agglomerative algorithm that ends up with a tree (dendrogram) that arranges the elements into a hierarchical structure \cite{mantegna2}.
%Therefore the use of the MST in the analysis of a financial market provides naturally also a hierarchical clustering over the market.
The filtering procedure linked to MST and SL has been succesfully applied to improve portfolio optimization \cite{cluster_portfolio}. 
 Another hierarchical clustering method, the Average Linkage (AL), has been shown to be associated to a slightly different version of spanning tree 
 \cite{spanning_tree_boot},  called  Average Linkage Minimum Spanning Tree. Another variant of Linkage methods, not associated to a spanning tree representation, 
 is the Complete Linkage (CL) \cite{cluster_book}. 
 
The MST is the first but not the only correlation-based filtered network studied in the literature.  
In particular the Planar Maximally Filtered Graph (PMFG) is a further step from the MST, that is able to retain a higher amount of information \cite{PMFG2,Tumminello05,MatlabPMFG}, having less strict topological constraint allowing to keep a larger number of links. The PMFG has been proven to have interesting practical applications, in particular in the field of investment strategies to hedge risk \cite{invest_periph}.

Since the MST has associated a clustering method, and the PMFG is a generalization of the MST, it could be raised the question whether the PMFG provides a clustering method that exploits this higher amount of information.
In a recent work \cite{DBHT} it has been shown that this is the case: the Directed Bubble Hierarchical Tree (DBHT) is a novel hierarchical clustering method that takes advantage of the topology of the PMFG yielding a clustering partition and an associated hierarchy. 
(For the DBHT algorithm refer to supplementary material of  \cite{DBHT} or, for a slightly modified version, to \cite{MatlabDBHT}.)
The approach is completely different from the agglomerative one adopted in the Linkage methods: the idea of DBHT is to use the hierarchy hidden in the topology of a PMFG, due to its property of being made of three-cliques 
\cite{DBHT,bubble}. In \cite{DBHT} the DBHT hierarchical clustering has been applied to synthetic and biological data, showing that it can outperform many other clustering methods. 
Since DBHT exploits the topology of the correlation network it can be viewed as an example of community detection algorithm in graphs \cite{fortunato_review}.

In this paper we present the first application of DBHT to financial data. To this purpose we have analysed the correlations among log-returns of $N = 342$ US stock prices,
across a period of 15 years (1997-2012). We have studied the structure of the clustering and we have compared the results with other clustering methods, namely: the  single Linkage, the average Linkage, the complete Linkage
and the k-medoids \cite{kmedoids} (a partitioning method strictly related to the k-means \cite{kmeans}).  
The perspective of our study focuses not only on the clusterings, but on the entire hierarchies associated to those clusterings, covering all the different levels of the hierarchical structures. We have also studied the dynamical evolution of these structures, describing how the hierarchies change with time. The dynamical perspective is crucial for applications, in particular for what concerns hedging risk and portfolio optimization.
For this reason we have given a particular attention to the effects of financial crises on the hierarchical structures, highlighting differences among the clustering methods.

 {\color{black} Another aspect we have focused on is the role of the market mode in shaping the clustering structure. 
To this aim we have carried out our analyses also on log-returns detrended of the market mode \cite{market_mode}: this procedure has been proven to provide a  more robust clustering  that can carry information not evident in the original correlation matrix \cite{market_mode}.
The comparison between the detrended and non-detrended correlation structures can shed light on the effect of common factors in the collective market dynamics especially during crisis.}

In order to compare quantitatively the amount of information retrieved by the different hierarchical clustering methods, we have taken the Industrial Classification Benchmark (ICB) as a benchmark community partition for the stocks and then we have compared it with the output of each clustering method. 
The idea is to use the degree of similarity between the ICB and the clustering as a proxy for the amount of information filtered by the methods. 
The degree of similarity is measured by using tools as the Adjusted Rand Index \cite{adj_rand} and the Hypergeometric hypothesis test \cite{enrichment}. This is not the first work comparing 
correlation-based clusterings and industrial sector classification; however to our knowledge the comparison has been performed only qualitatively so far \cite{sect_LSTE}, with the exception of ref. \cite{market_mode} where
 however only one clustering method is analysed. In Ref. \cite{kullback_measure}
the authors have compared quantitatively different methods in terms of amount of filtered information: yet this comparison was performed without looking at the industrial sector classification and by assuming a multivariate Gaussian distribution for the stocks returns \cite{mantegna2}. 
Our approach  is instead model-free. 
This is a relevant improvement since multivariate Gaussian models are known to be inaccurate to describe stocks returns \cite{stylized_facts,mantegna_book} and, more  generally, the correlation-based networks obtained from real data have been found to be incompatible with some widespread models for asset returns \cite{bonanno_mst}.

The paper is organized as follows. {\color{black} In Section ``Methods'' we describe the main tools we have used to carry out the analyses. In Section ``Dataset and preliminary analyses'' we present the dataset and some 
preliminary empirical analyses on it}. 
In Section ``Static analysis'' we perform a set of analyses on correlations and clusterings calculated by taking the whole 15 years period as time window,
hence ending up with only one hierarchical structure of dependences for each method. In particular we compare the
clustering compositions in terms of ICB supersectors for different clustering methods and we measure the similarity between clusterings and ICB partition. 
In Section ``Dynamical analysis'' we perform instead analyses using a dynamical approach, with moving time windows.
In Section ``Discussion'' we discuss the results and the future directions of our work. {\color{black} More details on the clustering methods we have analysed are reported in the Supplementary Information (SI)}.

\section*{Methods}
 {\color{black} To investigate and compare the DBHT with the other clustering methods we have performed a series of analyses that aim at describing different aspects of each clustering, from their structure to the economic information they contain. 
Here we introduce and describe the tools we have used.

\subsection*{Clustering methods}
In this paper we compare four different clustering methods, namely: Single Linkage (SL), Average Linkage (AL), Complete Linkage (CL), Directed Bubble Hierarchical Tree (DBHT), k-medoids.
All these methods require a suitable distance measure and DBHT also requires an associated similarity measure. 
In this paper we use the Pearson's correlation coefficients $\rho_{ij}$ between pairs of stocks as similarity measure and we use the associated Euclidean distance $D_{ij}=\sqrt{2(1-\rho_{ij})}$; $\rho$ and $D$ are $N\times N$ symmetric matrices with, respectively, all ones and all zeros on the diagonal. 
Let us here briefly list the main features and requirements of these clustering methods.
 
 \begin{itemize}
  \item \textbf{Single Linkage} (SL), is a hierarchical clustering algorithm \cite{cluster_book}. 
  Given the distance matrix, it starts assigning to each object its own cluster, and then at each step merges together the least distant pair of clusters, until only one cluster remains. 
 At every step the distance between clusters $A$ and $B$ is updated taking the minimum distance between elements inside the clusters: 
 %$d_{A,B}=\min_{a \in A, b \in B} D(a,b)$ with $a\in A$ and $b\in B$.
  \begin{equation}
  \label{eq:SL}
   d_{A,B}=\min_{a \in A, b \in B} D(a,b) 
  \end{equation}
SL is called an \emph{agglomerative} clustering, since it begins with a partition of $N$ clusters and then proceeds merging them. 
The final output is a dendrogram, that is a tree showing the hierarchical structure. 
The distance measure  in this dendrogram is an ultrametric distance (see for instance \cite{mantegna1}). 
A cluster partition can be obtained by choosing the number of clusters (that is therefore a free parameter) and cutting the dendrogram at the appropriate level. 
The SL algorithm is strictly related to the one that provides a Minimum Spanning Tree (MST) 
 \cite{mantegna2}. 
The MST is a tree graph, it contains exactly $N-1$ links.
There is therefore a strict relation between the two tools. 
However the MST retains some information that the SL dendrogram throws away \cite{mantegna2} .
%, and it has been used as topological tool in Econophysics since the work of Mantegna \cite{mantegna1}. 
%The MST can be generated starting with an empty graph:  after sorting all the correlations in $C$ in descending order, add a weighted link between the two stocks/nodes with the highest correlation, and then go ahead with the next highest pair correlation; whenever the new link to add  generates a loop, do not add that link and skip to the next one, until all the list is checked. 
 %This tree contains exactly $N-1$ links.  It can be shown \cite{mantegna2} that the MST algorithm is basically the SL procedure carried out until the graph is completely connected. 
% There is therefore a strict relation between the two tools. However the MST retains some information that the SL dendrogram throws away \cite{mantegna2} .
%   (MST) \cite{mantegna1}, that is a tree $MST(t_k)$ ($MST^{R}(t_k)$) associated to each given distance matrix $D(t_k)$ ($D^R(t_k)$). Each tree contains exactly $N-1$ links.
%   To each MST is naturally associated a dendrogram, generated by the linkage algorithm that runs in parallel with the MST construction . 

  \item \textbf{Average Linkage} (AL) \cite{cluster_book} is a hierarchical clustering algorithm which is constructed in the same way as the SL, but with Eq. \ref{eq:SL} replaced by:
   \begin{equation}
   d_{A,B}= \mathop{mean}_{a \in A, b \in B} D(a,b) 
  \end{equation}
  
  \item \textbf{Complete Linkage} (CL) \cite{cluster_book} is another variant of SL, where Eq. \ref{eq:SL} is replaced by:
    \begin{equation}
   d_{A,B}=\max_{a \in A, b \in B} D(a,b) 
  \end{equation}
  
  \item \textbf{Directed Bubble Hierarchical Tree} (DBHT) \cite{DBHT}, is instead a novel hierarchical clustering method that exploits the topological property of the PMFG (Planar Maximally Filtered Graph)  in order to assign the clustering.
 The PMFG is a generalization of the MST, that is included in the PMFG as a subgraph. 
 It can be constructed by recursively joining together with an edge nodes with the smallest distance restricting the procedure to only edges that do not violate graph planarity condition.
The PMFG contains the MST as a sub-graph but it retains a larger number of links than the MST ($3(N-2)$ instead of $N-1$).
%  In particular it can be shown that each PMFG contains exactly $3(N-2)$ links.
 The basic elements of a PMFG are three-cliques (subgraphs made of three nodes all reciprocally connected). 
 The DBHT exploits this topological structure, and in particular the distinction between separating and non-separating three-cliques, to identify a clustering partition of all nodes in the PMFG \cite{DBHT}. 
  
The Linkage algorithms look at the sorted list of distances $D_{ij}$, build the dendrogram by gathering subsets of stocks with lowest distances and then find the clustering from the dendrogram after choosing the number of clusters as free parameter.  
 The DBHT instead reverses this order: first of all the clusters are identified by means of topological considerations on the planar graph, then the hierarchy is constructed both inter-clusters and intra-clusters.  
The difference involves therefore both the kind of information exploited and the methodological approach.
    
  \item \textbf{k-medoids} is a partitioning clustering method closely related to k-means \cite{kmeans}. 
  It takes the number of clusters $N_{cl}$ as an input. 
  The algorithm is the so called Partitioning Around Medoids (PAM), and is constructed in the following steps:
  1) select randomly $N_{cl}$ ``medoids'' among the $N$ elements; 
  2) assign each element to the closest medoid; 
  3) for each medoid, replace the medoid with each point assigned to it and calculate the cost of each configuration;
  4) choose the configuration with the lowest cost;
  5) repeat 2)-4) until no changes occur.
%        \begin{enumerate}
%         \item select randomly $N_{cl}$ ``medoids'' among the $N$ elements;
%         \item assign each element to the closest medoid;
%         \item for each medoid, replace the medoid with each point assigned to it and calculate the cost of each configuration;
%         \item choose the configuration with the lowest cost;
%         \item repeat 2)-4) until no change occurs.
%        \end{enumerate}
  This method, differently from previous, is not a hierarchical method and therefore does not provide a dendrogram but only a partition.
 \end{itemize}

}

\subsection*{Measuring clustering heterogeneity: disparity}
\label{subsec:disp_def}
 {\color{black} 
%A clustering feature that we found worth analysing, especially in relation to the role of market mode in shaping the correlation based community, is the distribution of clusters size. 
Different clustering methods can differ greatly in the clustering output even under the same input conditions. 
Some clustering methods might provide a subdivision into a few very large clusters and many small clusters, whereas other might display a more homogeneous distribution. 
In order to characterize the distribution of the cluster cardinality with a single quantity we have calculated for each clustering partition the coefficient of variation:}
\begin{equation}
  y = \frac{\sigma_S}{\langle S \rangle},
  \label{eq:disparity}
 \end{equation}
where $\sigma_S$ is the standard deviation 
 \begin{equation}  
  \sigma_S =\sqrt{\frac{1}{N_{cl}-1} \sum_a ( S_a-\langle S \rangle)^2},
 \end{equation}
and the normalization factor $\langle S \rangle$ is the average
 \begin{equation}
    \langle S \rangle = \frac{1}{N_{cl}}\sum_a S_a,
 \end{equation} 
with $S_a$ being the cardinality  (number of stocks in) cluster $a$ and $N_{cl}$ the number of clusters. 
In the limit of homogeneous arrangement of stocks among the clusters (i.e. each cluster has the same number of stocks), we obtain $\sigma_S = 0$ and then $y=0$. The 
 higher is the degree of heterogeneity in the distribution of sizes, the higher is $\sigma_S$ and therefore $y$.
 {\color{black} In the following we have used the expression ``disparity'' to refer to $y$, in order to stress the fact that  we use it as a measure of heterogeneity in clusters' sizes.}

\subsection*{Measuring clustering similarity: Adjusted Rand Index}
\label{appendix:ari_def}
 {\color{black} To measure the amount of economic information  in each correlation based clustering} we have used the Adjusted Rand Index ($\mathcal{R}_{adj}$) \cite{adj_rand} which is a tool conceived to compare different clusterings on the same set of items \cite{comparing_clustering}.
An industrial sector classification is indeed nothing but a partition in communities of the $N$ stocks. {\color{black} Therefore we can take the similarity between clustering and industrial sector classification as a proxy for the 
information detected by the clustering method}.

In particular, given two clusterings on the same set of items, $\mathcal{R}_{adj}$ returns a numerical value equal to 1 for identical clusterings and to 0 for clusterings completely independent which are undistinguishable from a random choice. 

The idea behind this measure is to calculate the number of pairs of objects that are in the same cluster in both clusterings, and then to compare this number with the one expected under the hypothesis of independent clusterings.
Specifically, and following the notation of \cite{comparing_clustering}, let us call $X$ the set of the $N$ objects (stocks, in our case). 
Let us call $Y=\{Y_1,...,Y_k\}$ a clustering which is a partition of $X$ into communities which are non-empty disjoint subsets of $X$ such that their union equals $X$:  $X=Y_1 \cup...\cup Y_k$ \cite{comparing_clustering}.
% or simply a clustering, that is ``a set $Y=\{Y_1,...,Y_k\}$ of non-empty disjoint subsets of $X$ such that their union equals $X$''
Let us also consider another different clustering $Y'$, containing $l$ clusters.
We call ``contingency table'' the matrix $M=\{m_{ij}\}$ with coefficients  
 \begin{equation}
  m_{ij} \equiv |Y_i \cap {Y'}_j|,
 \end{equation}
 i.e. the number of objects in the intersection of clusters $Y_i$ and
 ${Y'}_j$ . 
 Let us call $a$ the number of pairs of objects that are in the same cluster both in $Y$ and in $Y'$, and $b$ the number of pairs that are in two different clusters in both $Y$ and $Y'$.
 Then the Rand Index is defined as the sum of $a$ and $b$, normalized by the total number of pairs in $X$:
 \begin{equation}
  \mathcal{R}(Y,Y') \equiv \frac{2(a+b)}{N(N-1)} = \sum_{i=1}^k \sum_{j=1}^l \binom{m_{ij}}{2}.
 \end{equation}
We then use, as null hypothesis associated to two independent clusterings, the generalized hypergeometric distribution and define the Adjusted Rand Index as the difference between the  Rand Index and its mean value under the null hypothesis, normalized by the maximum that this difference can reach:
\begin{equation}
\label{eq:ari}
 \mathcal{R}_{adj}(Y,Y') \equiv \frac{\sum_{i=1}^k \sum_{j=1}^l \binom{m_{ij}}{2}-t_3}{\frac{1}{2}(t_1+t_2)-t_3},
\end{equation}
where
\begin{equation}
 t_1=\sum_i^k \binom{|Y_i|}{2} ~~ , ~~ t_2=\sum_j^l \binom{|{Y'}_j|}{2} ~~,~~ t_3=\frac{2t_1t_2}{N(N-1)}.
\end{equation}
We have $\mathcal{R}_{adj} \in [-1, 1]$ , with $1$ correspondent to the case of identical clusterings and $0$ to two completely un-correlated clusterings. 
Negative values instead show anti-correlation between $Y$ and $Y'$ (that is, the number of pairs classified in the same way by $Y$ and $Y'$ is less  than what expected  {\color{black} assuming} a random overlapping between the two clusterings).

\subsection*{Hypergeometric test for cluster-industry overexpression}
\label{appendix:hypergeometric_test}
The Adjusted Rand Index provides an overall measure of similarity between the clustering partition and the industrial classification. 
In order to analize how much each industrial sector is retrieved by each cluster we must look at the stocks respectively within a given cluster and a given sector and measure the number of stocks in common.
If the  percentage of stocks in common  between a cluster and an industrial sector is sensitively higher than what expected from a random overlapping of communities we say that the cluster \emph{overexpresses} a specific sector.
To quantify such overexpression we use a statistical one-tail hypothesis test, where the null hypothesis is the Hypergeometric distribution which describes the probability that by random chance two communities of given sizes have in common $k$ objects over a total of $N$ \cite{feller2008introduction,bipartite_net}. 
%the number of objects in common between two overlapping and independent communities is equal to \cite{bipartite_net}. 
%This distribution is a  generalization of the Binomial distribution and describes the probability $P(k)$ that the number of objects in common between two overlapping and independent communities is equal to $k$, taking into account the size of the two communities and the overall number of objects.
In particular, let us call $Y_i$ a cluster in our clustering and $Y'_j$ a sector. 
We want to verify whether $Y_i$ overexpresses $Y'_j$. 
If $k$ is the number of stocks in common between $Y'_j$ and $Y_i$, and $|Y_i|$ , $|Y'_j|$ are the cardinalities of the cluster and the sector respectively, then  
the Hypergeometric distribution is \cite{feller2008introduction}:
\begin{equation}
\label{eq:hypog_test}
 P(X=k) ~=~ \frac{{|Y'_j|\choose k}{N-|Y'_j|\choose|Y_i|- k }}{{N\choose|Y_i| }}. 
\end{equation}
 This is the null hypothesis for the test: to be distinguishable by a random overlap the number $k$ of stocks in common must be significantly different from a random overlap and therefore $P(X=k)$ must be small.
 If $P(X=k)$ is less than the significance level, then it is said that the test is rejected. 
 If the test is not rejected, then it means that we cannot reject the hypothesis that the $k$ stocks in $Y_i$ coming from a sector $Y'_j$ are picked up just by chance, without any preference for that sector.
 If instead the test is rejected, we conclude that the cluster $Y_i$ overexpresses the sector $Y'_j$. 
 We have chosen a significance level of $1\%$, together with the Bonferroni correction for multiple tests, which reduces remarkably the significance level of each test \cite{feller2008introduction} (more details in Section ``Industries overexpression'').

% Results and Discussion can be combined.
%\section*{Results}

\section*{Dataset and preliminary analyses}
\label{sec:dataset}

The correlation structure studied in this paper concerns $N=$ 342 stocks from the New York Stock Exchange (NYSE). 
A complete description of the dataset is in Supplementary Information (SI). 
%\ref{appendix:dataset}
We have analysed the closing 
daily prices $P_i(t)$ with $i=1,...,N$, during the time between 1 January 1997 to 31 December 2012
(4026 trading days). From the prices, we have calculated the daily log-returns \cite{stylized_facts,mantegna_book}: 
 
  \begin{equation}
   r_i(t) \equiv \log(P_i(t))- \log(P_i(t-1)).  
 \end{equation}
 
 From the set of $N$ log-return time series over a time window $T=[t_{start},t_{end}]$ we have then calculated the $N \times N$ correlation matrix $\rho(T)$, whose elements are given by the Pearson estimator \cite{pearson}:
  \begin{equation}
  \label{eq:correlation}
  \rho_{ij}(T) = \frac{\langle r_i(t) r_j(t) \rangle_{T}}{\sqrt{[\langle r^2_i(t) \rangle_{T} - \langle r_i(t) \rangle_{T}^2][\langle r^2_j(t) \rangle_{T} - \langle r_j(t) \rangle_{T}^2] }},
 \end{equation}
 where $\langle ... \rangle_{T}$ represents the average over the  time window $T$.
 The clustering analysis is then carried out on the distance matrix D, with elements $D_{ij}(T)=\sqrt{2(1-\rho_{ij}(T))}$. 
For the analysis on moving windows we used an exponentially smoothed version of the Pearson estimator in Eq.\ref{eq:correlation}, where terms in the average are multiplied by a weight $w_t=w_0\exp(\frac{t-t_{end}}{\theta})$ with $t \in T$ according to their  temporal distance from the last trading time $t_{end}$ in the window $T$.
 This exponential smoothing scheme \cite{exp_smoothing} allows to mitigate excessive sensitiveness to outliers in remote observations. 
 The parameter $\theta$ has been set to $\theta = T / 3$ according to previously established criteria \cite{exp_smoothing}.
 
 \begin{figure}[ht!]
 \begin{center}
\begin{tabular}{llc}
\hspace{-3em}
  \includegraphics[scale=0.9]{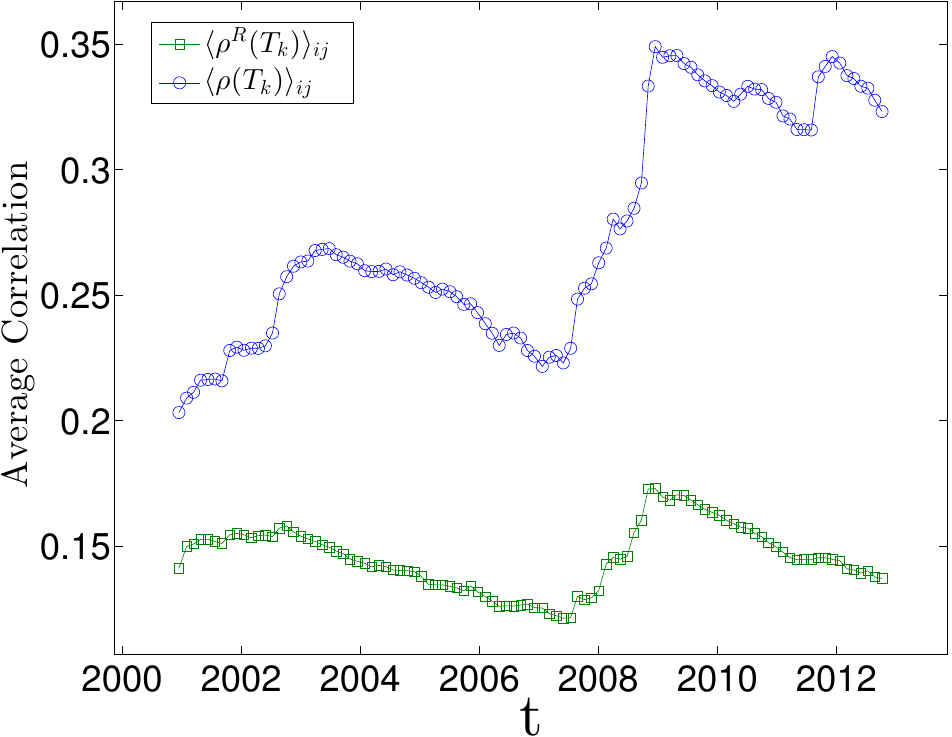} 
\end{tabular}
\end{center}
 \caption{\label{fig:avg_corr_vol} {\bf Demonstration that the average correlation evolves during time with large changes during periods of market instability}. 
 The figure reports the average correlation for each time window $T_k$ with $k = 1, ..., n$ ($n=100$, each time window has length $L=1000$ trading days), for both non-detrended (blue circles) and detrended log-returns (green squares). 
 The average correlation is highly reduced by detrending the market mode.} 
\end{figure}
 
 By using this moving time window approach we have performed a set of preliminary analyses on the average correlation of our set of stocks, looking in particular at the 2007-2008 financial crisis. Specifically, we have considered a set of $n=100$ overlapped time windows $T_k$ ($k=1,...,n$) of length $L=1000$ trading days (four years) with 30 days shift between subsequent windows.  
 The average correlation $\langle \rho(T_k) \rangle_{ij}$ on these windows  is shown in Fig.~\ref{fig:avg_corr_vol} (blue circles) for $L=1000$ and $n=100$.  
 To test robustness, we have verified that the results are similar also for other window sizes, namely $L=750$ and $L=1250$.
 
 We have also studied the detrended log-returns, i.e. log-returns subtracted of the average return over all the stocks. 
 %, and we have calculated the Pearson's correlations on it.
 Specifically, following \cite{market_mode}, we have considered a single factor model for each stock $i$:
   \begin{equation}
    r_i(t) = \alpha_i + \beta_i I(t) +c_i(t),
    \label{eq:market_mode}
   \end{equation}
 where the common market factor $I(t)$ is the market average return, $I(t)=  \frac{1}{N} \sum_ir_i(t)$ and the residuals, $c_i(t)$, are the log-returns detrended by the market mode.
 After estimating the coefficients $\alpha_i$ and $\beta_i$ with a linear regression, the residuals $c_i(t)$ can be calculated and used to evaluate the new correlation matrix \cite{market_mode}. 
 We denote this matrix, estimated in the time window $T_k$ with $\rho^R(T_k)$. 
 We refer to the analyses  based on this kind of correlation matrix as the ``detrended case''.
 These detrended correlation matrices are worth analyzing since they have been found to provide a richer and more robust clustering \cite{market_mode} that can carry information not evident in the original  correlation matrix \cite{multifactor_clustering}. 
 In this paper we have carried the analyses using both detrended and non-detrended log-returns, compared the two and looked for differences that might highlight the effect of common factors on the market correlation structure. 
 %{\color{black} For these reasons in the rest of the paper we will always show results on detrended log-returns, displaying the correspondent non-detrended analyses only when the comparison  reveals interesting features}.
  
  In Fig. \ref{fig:avg_corr_vol} it is shown the average correlation for these detrended correlation matrices, i.e. $\langle \rho^R(T_k) \rangle_{ij}$, compared to the average correlation for the non-detrended correlation matrices $\langle \rho(T_k) \rangle_{ij}$. 
  As one can see, the subtraction of the market mode decreases by about 
  $50\%$ the average level of correlation, pointing out the important role of the market factor in the correlation structure. However, we can still observe the increase correspondent to the {\color{black}  financial crisis in 2007-2008}. Moreover, and   interestingly, the level of correlation reduces after a peak in 2009, unlike the non-detrended case. This fact suggests that, although the market mode plays an important role in terms of average amount of correlation, yet 
  the peak of the {\color{black} last financial crisis} seems not to be only a global market trend. {\color{black} We therefore suggest that it could involve}, to some extent, the internal dynamics among stocks that remains after the subtraction. 

\section*{Results}
\subsection*{Static analysis}
\label{sec:static_analysis}

\begin{figure}[ht!]
\begin{center}
\begin{tabular}{l}
  \includegraphics[scale=0.9]{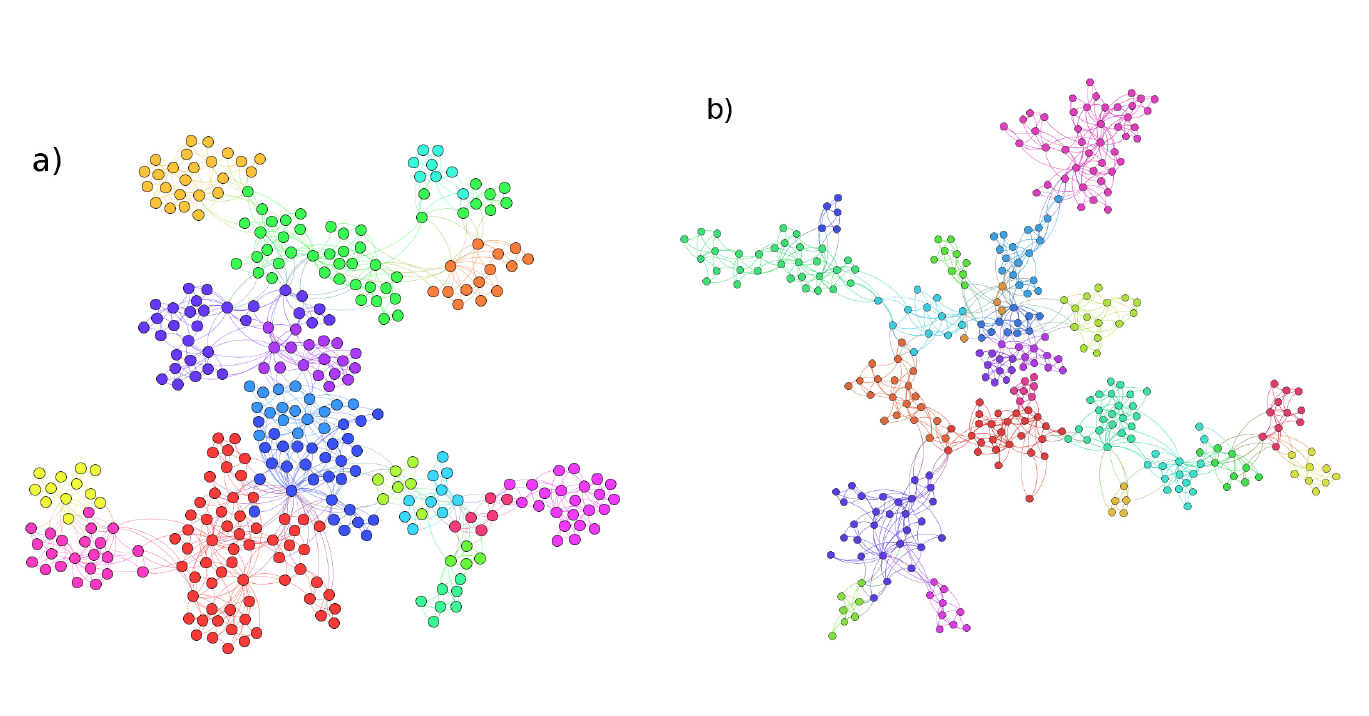}
\end{tabular}
\end{center}
 \caption{\label{fig:pmfg} {\bf Visualization of the Planar Maximally Filtered Graph (PMFG) and DBHT clusters, for both non-detrended and detrended log-returns}.
 a) PMFG calculated on the entire period 1997-2012, using non-detrended log-returns. Stocks of the same color belong to the same DBHT cluster.
 b) PMFG calculated on the same data as in a), but using detrended log-returns. Stocks of the same color belong to the same DBHT cluster.} 
\end{figure}

 \subsubsection*{DBHT clusters composition}
\label{appendix:histo_cl}

\begin{figure}[ht!]
\begin{center}
\begin{tabular}{l}
  \includegraphics[scale=0.9]{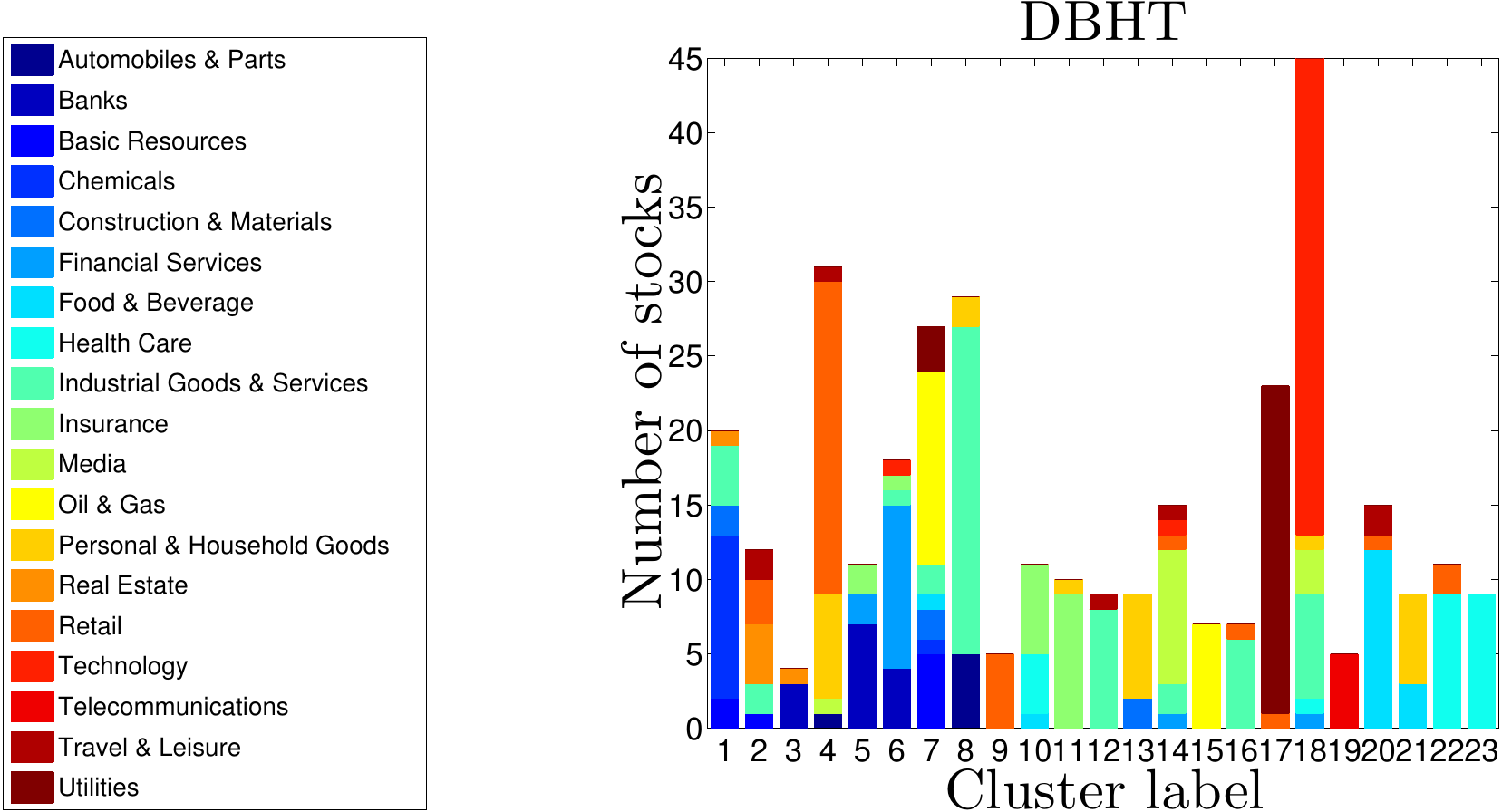}
\end{tabular}
\end{center}
 \caption{\label{fig:histo_cl_composition_dbht_detrended} {\bf Composition of DBHT clusters in terms of ICB supersectors}.The x-axis represents the cluster labels, the y-axis the number of stocks in each cluster. 
 Each colour corresponds to an ICB supersector (legend on the left hand side). The clustering is obtained by using log-returns detrended by removing the market mode. See the Fig. S3 in S1 file
 for the case non-detrended.} 
\end{figure} 

In this section we present results for the PMFG and DBHT clustering method applied to the set of data described in the previous section. In particular we have computed the PMFG and the correspondent DBHT 
clustering in the time period 
from 1997 to 2012 and we plot it in Fig. \ref{fig:pmfg} a) where we highlight, with the same color, stocks belonging to the same DBHT cluster. In the same figure (Fig. \ref{fig:pmfg} b)) we plot the PMFG calculated by using the 
detrended log-returns (Eq. \ref{eq:market_mode}) as comparison. This PMFG looks more structured than the first one, with more homogeneous clustering sizes. 

We have then analysed the DBHT clustering structure in terms of industrial sectors. 
It is well known that the hierarchical structure of the stock return correlations shows a deep similarity with the industrial sectors categorization \cite{mantegna1} \cite{sect_LSTE} \cite{multifactor_clustering}. 
This fact supports the intuitive  
argument that returns of stocks in the same industrial sector are affected mainly by the same flows of information and economic enviroment. 
We can turn around the reasoning and claim that thus a desirable feature of a clustering method on stocks data is to retrieve, to some extent, the industrial sector classification.
% {\color{black} Since the subtraction of market mode has be proven to provide a clustering that is more robust and more similar to the industrial classification \cite{market_mode},
% we will focus on log-returns detrended of the market mode. In SI we describe the same analyses for undetrended log-returns.}
We will refer to the 
 Industrial Classification Benchmark (ICB) \cite{ICB}; this categorization divides stocks into 19 different supersectors, that in turn are gathered in 10 different Industries. For more details on the composition 
 of our dataset in terms of ICB supersectors, refer to SI. {\color{black} Let us point out that we have run all the analyses in this paper also by using the Yahoo industrial partitioning, obtaining similar results}.
 
  In Fig. \ref{fig:histo_cl_composition_dbht_detrended} we report a graphical summary of the clusters obtained applying the DBHT method to the whole time window of data (1997-2012), by using detrended log-returns (the clustering shown in Fig. \ref{fig:pmfg} b)). In S1 file we describe the same analyses for undetrended log-returns.
  
  The DBHT retrieves a number of clusters, $N_{cl}$, equal to {\color{black} $23$}: to each cluster is associated a bar, whose height represents the number of stocks in the cluster. 
  Each bar is made of different colours, showing the composition of each cluster in terms of 
  ICB supersectors. The legend on the left of the graph reports the corresponding industrial supersectors. 
  Please note that the colours in Fig. \ref{fig:histo_cl_composition_dbht_detrended} identify the ICB supersectors, and they  have nothing to do with colours in Fig. \ref{fig:pmfg}, that identify DBHT clusters.
  
  {\color{black} The largest cluster contains 45 stocks ($13\%$ of total), the smallest 4. The average size is $14.8$. As we can see, there are several supersectors that are overexpressed by one or more clusters, either alone or together 
  with other supersectors: Oil \& Gas (clusters 7 and 15), Technology (cluster 18), Utilities (cluster 17), Retail (cluster 9), Health Care (clusters 22 and 23), Food \& Beverage (clusters 20 and 21), Personal \& Household Goods
  (cluster 21), Industrial Goods \& Services (clusters 12 and 16), Insurance (cluster 11) and Telecommunications (cluster 19). Moreover, there are clusters that, although showing a mixed
  composition, are composed by supersectors strictly related: clusters 5 and 6 are made of Banks, Financial Services and Insurance, all supersectors that the ICB gathers in the same industry (Financial) at the superior hierarchical
  step. Similarly, cluster 21 is made entirely of Food \& Beverage and Personal \& Household Goods stocks, that are both classified in Consumer Goods industry.}

\subsubsection*{Other clustering compositions}
 \label{sec:linkage_clusters}
 
 \begin{figure}[ht!]
\begin{center}
\begin{tabular}{l}
  \includegraphics[scale=0.9]{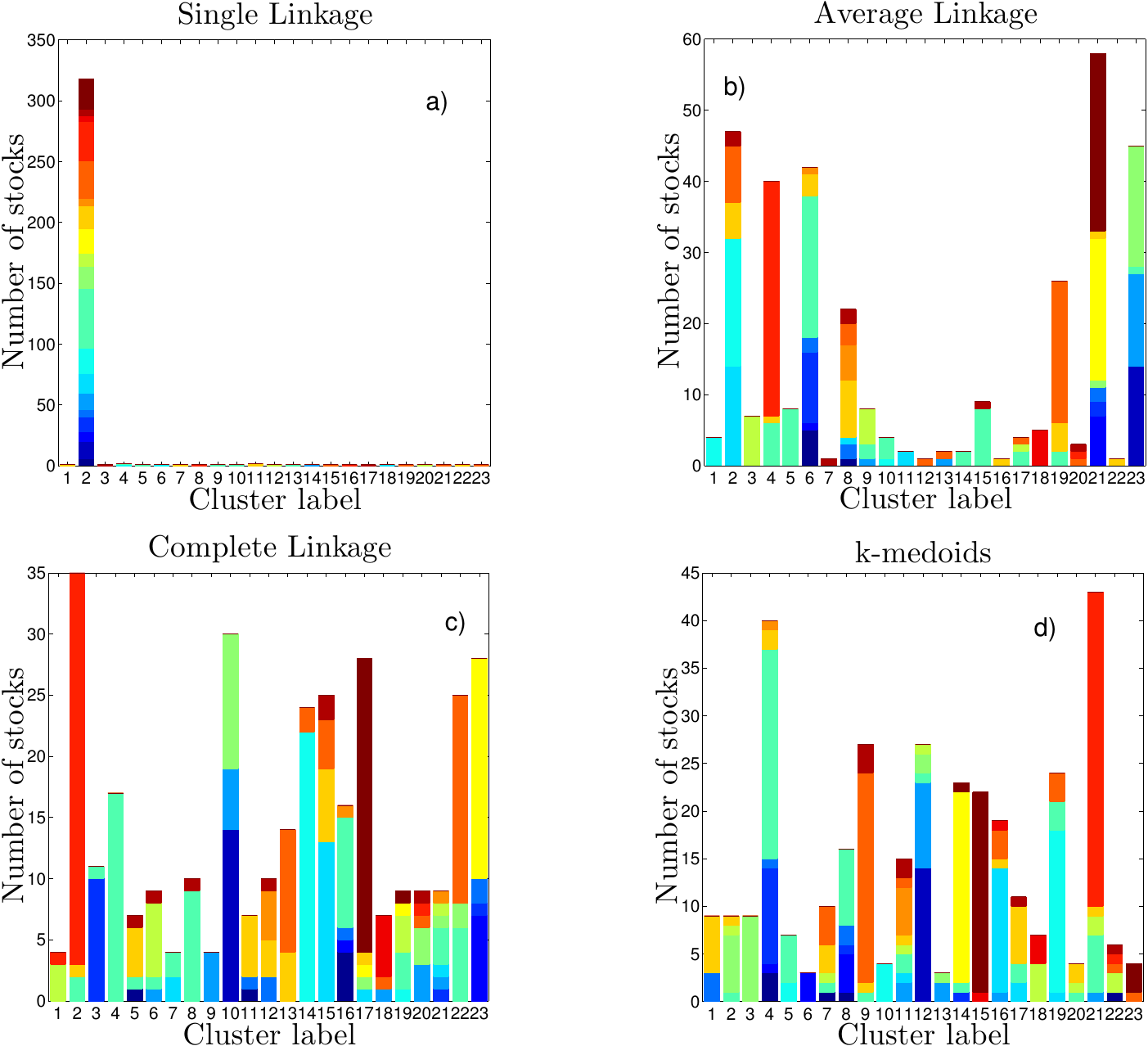}
\end{tabular}
\end{center}
 \caption{\label{fig:histo_cl_composition_linkage_detrended} {\bf Composition of clustering in terms of ICB supersectors}. The x-axis represents the  cluster labels, the y-axis the number of stocks in each cluster. 
 Each colour corresponds to an ICB supersector (the legend is the same as in Fig. \ref{fig:histo_cl_composition_dbht_detrended}). The graphs show the results for a) SL clustering, b) for AL, c) for CL and d) for k-medoids.
 The clustering is obtained by using log-returns detrended by removing the market mode.} 
\end{figure}

We have applied other clustering methods on the same data and compared results with DBHT clustering. 
The clustering methods considered are Single Linkage (SL), Average Linkage (AL), Complete Linkage (CL) and k-medoids. 
The number of clusters, that unlike the DBHT is a free-parameter for these methods, has been chosen equal to {\color{black} $23$} in this case, in order to compare the bar graphs with the
Fig. \ref{fig:histo_cl_composition_dbht_detrended} for DBHT.
We plot in Fig. \ref{fig:histo_cl_composition_linkage_detrended} a), b), c) and d) the clusters composition obtained by using these four clustering methods, namely SL, AL, CL and k-medoids. As for DBHT, the same analyses by using 
non-detrended log-returns are discussed in SI.
 
 {\color{black} First of all we can observe that for SL there is a strong heterogeneity in the size of clusters, with the presence of a giant cluster containing 318 stocks, and the other clusters made of one, two or three stocks only. 
 This giant cluster contains stocks of all ICB sectors.} 
 
 {\color{black} The AL case shows instead a more structured clustering: the size of the largest cluster
  shrinks to 58 stocks, and 6 different clusters of medium size (20-40 stocks) appear. Moreover, these clusters show a much higher overexpression of
  %(that is, a percentage significatly higher than what expected by change of) 
  supersectors than SL, such as Technology (cluster 4),
  Industrial Goods \& Services (cluster 5 and 15), Media (cluster 3) and Financial related supersectors (cluster 23). However there are still 10 clusters whose size is at most 4 stocks.}
 
 {\color{black} For the CL and the k-medoids the supersectors overexpression is further improved, becoming as rich as the DBHT one. 
 Especially CL shows  overexpression of Technology (cluster 2), Industrial Goods \& Services (clusters 4 and 8),
 Utilities (cluster 17), Oil \& Gas (cluster 23), Health Care (cluster 14) and Financial Services (cluster 9). 
 Similar overexpressions are found for the k-medoids case.}
 
 These first comparisons are however made under a specific choice of the number of clusters ({\color{black} $23$}), given by the DBHT. One could wonder what happens changing this parameter, i.e. moving along the hierarchical structure
 provided by each clustering method.
 Let us stress that the DBHT method gives automatically the number of clusters that is instead an adjustable parameter for the other methods. However, DBHT can also be analysed for a varying number of clusters by thresholding 
 over the clustering hierarchical structure. 
 In the following Sections we discuss a set of quantitative analyses that explore
 all the hierarchical levels of the DBHT and the other clustering methods. 

\subsubsection*{Disparity in the clusters size}
\label{subsec:disparity}

\begin{figure}[ht!]
 \begin{center}
\begin{tabular}{l}
\hspace{-3em}
  \includegraphics[scale=0.9]{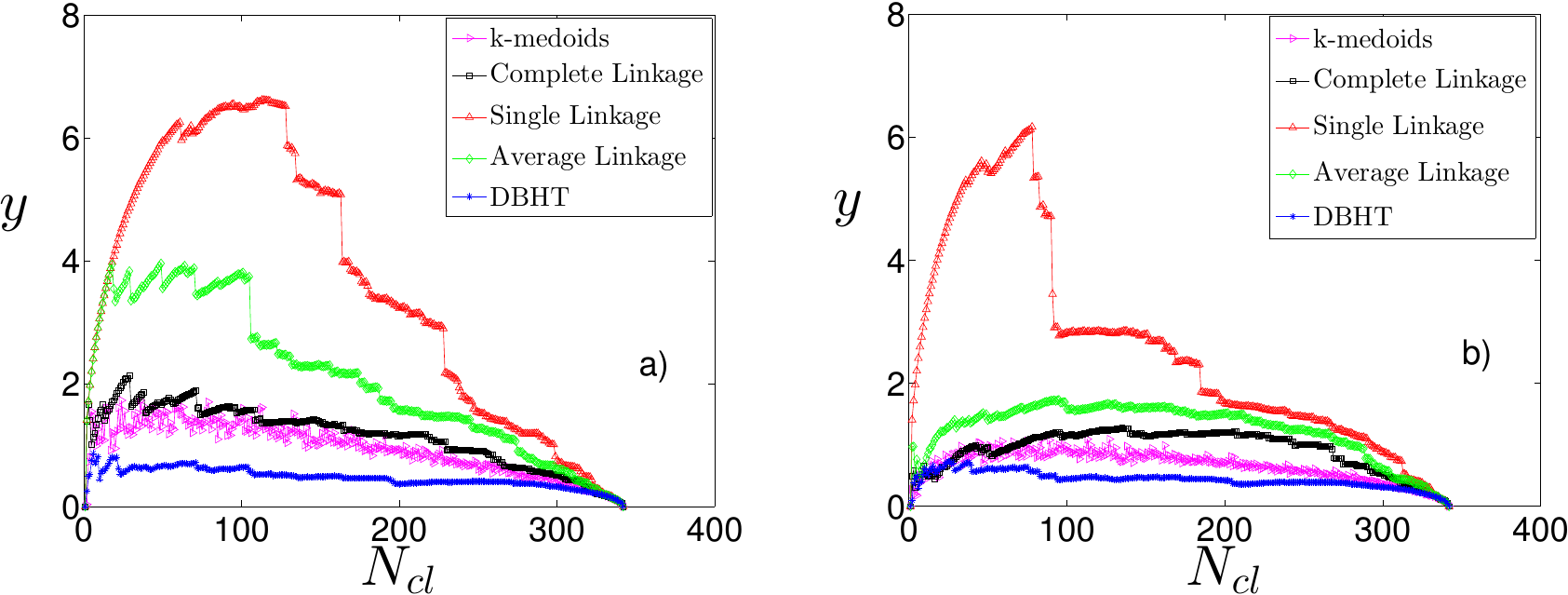}
\end{tabular}
\end{center}
 \caption{\label{fig:disp_ncl} {\bf Demonstration that different clustering methods show different degrees of disparity in the clustering structure}. The disparity measure $y$ is shown for clusterings
 at different hierachical levels as function of $N_{cl}$ in the dendrograms, for a) non-detrended log-returns and b) detrended log-returns.} 
\end{figure} 

In the previous Section we have seen that the SL shows a giant cluster that contains more than $90\%$ of stocks, whereas DBHT, CL and k-medoids methods have a more homogenous distribution of cluster sizes and the AL seems to be an intermediate case.

Let us here check whether this difference in the structure depends 
 on the choice of the number of clusters for the linkage methods, which might be penalising the {\color{black} SL with respect to the others}.  
 In order to do that, we vary the number of clusters $N_{cl}$ for each clustering method by cutting the dendrograms at different levels. For the k-medoids, for which no dendrogram is present, $N_{cl}$ is simply an input parameter of the algorithm. 
 We then calculate the measure of disparity introduced in Methods (Eq. \ref{eq:disparity}).
 
In Fig. \ref{fig:disp_ncl} we show, for each clustering method, how the disparity measure varies with $N_{cl}$. The graph a) shows the non-detrended case, the graph b) the detrended case. 
As we can see the SL provides the higher disparity in both cases, regardless of $N_{cl}$, then the AL, CL and k-medoids follow.
 The DBHT values are below all of them, this means that the DBHT clustering provides a more {\color{black} homogeneous} community assignment at any level of the correlation hierarchy.
  Moreover, in the market mode case the SL and the AL show the highest values of disparity for $N_{cl}$ in the interval 50-100. The CL and DBHT have instead a flatter pattern, 
 with the highest values occurring for lower values of $N_{cl}$. Looking at the detrended case in Fig. \ref{fig:disp_ncl} b), the removal of the market mode smooths also the pattern of the AL, whereas the SL is even sharper.
 Overall, subtracting the market mode makes the clusterings more homogeneous, suggesting that the largest clusters that emerged in SL and AL in the non-detrended case are associated to the market mode dynamics. 
 
 {\color{black} The algorithms of SL and
  AL are indeed expected to be more sensitive to the market mode. In the iterative procedure that generates the SL dendrogram, for instance, the correlation between two new clusters is defined as the maximum correlation between   elements of the first cluster and elements of the second one: since the most part of correlation in the market is due to the market mode \cite{market_mode} such an algorithm is likely to force many clusters to join  the cluster made of the most influencial stocks in the market, resulting in a giant cluster and high disparity. 
  The AL is less sensitive to this effect as the inter-clusters correlation is defined as the average of correlations;  for the CL the minimum correlation is chosen,  resulting  -unsurprisingly- in the lowest value of disparity. For what concerns the DBHT it is probably the topology of PMFG, which is more structured and clusterised than the MST,  to provide a lower sensitivity to the market mode dynamics.}
 
 We can conclude that, from the point of view of the disparity measure, the analysed clustering methods provide quite different structures at any level of the dendrograms.
 The DBHT yields the most homogeneous clustering, whereas the SL displays the highest levels of disparity. 
 
 \label{subsec:disparity_static}
 \label{subsec:linkage_clusters}

\subsubsection*{Retrieving the industrial sectors}
\label{subsec:ind_sectors}

\begin{figure}[ht!]
\begin{center}
\begin{tabular}{l}
\hspace{-3em}
  \includegraphics[scale=1.2]{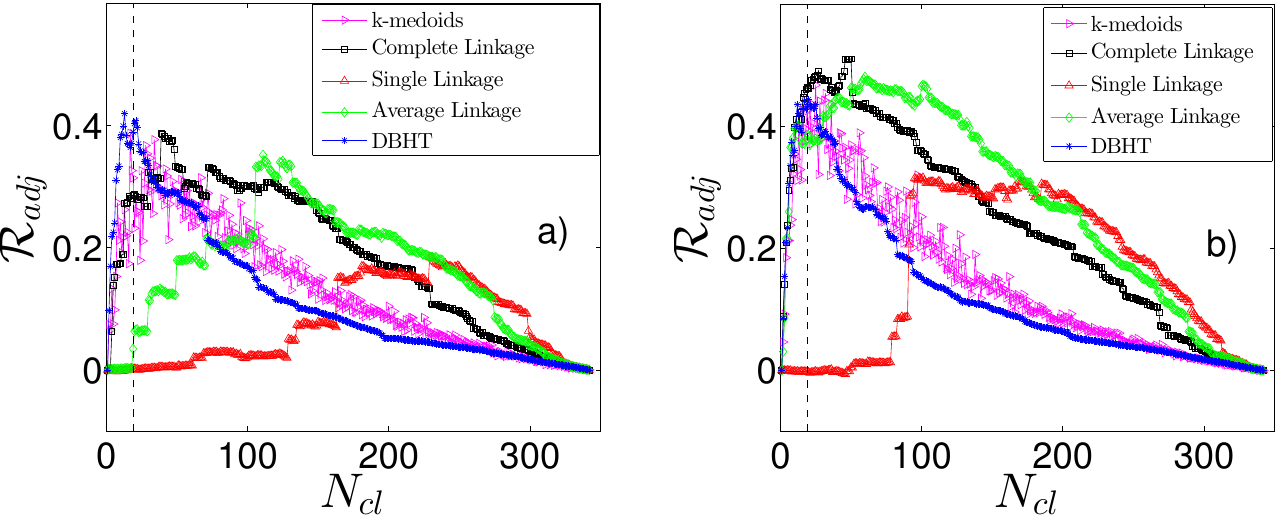}
\end{tabular}
\end{center}
 \caption{\label{fig:ari_sect_ncl} {\bf Demonstration that different clustering methods retrieve different amount of industrial sector information}. The Adjusted Rand Index $\mathcal{R}_{adj}$ between clustering and ICB supersectors 
 is shown for different number of clusters $N_{cl}$. In a) 
 correlations are calculated on non-detrended log-returns, in b) are calculated on detrended log-returns. The vertical dashed line shows the value ($N_{cl}=19$) correspondent to the actual number of ICB supersectors.} 
\end{figure}

In this Section we quantify the similarity between clustering and ICB, by varying the number of clusters $N_{cl}$. We have used the Adjusted Rand Index introduced in Methods (Eq. \ref{eq:ari}).
 In Fig. \ref{fig:ari_sect_ncl} we show the results of the first set of analyses.
 We have applied the five clustering methods to the entire time-period 1997-2012 and then we have varied the number of clusters $N_{cl}$. 
For each clustering obtained in this way we have calculated the Adjusted Rand Index $\mathcal{R}_{adj}$ between that clustering and the 
ICB supersector partition. In this way we have explored the entire hierarchical structure.

Fig. \ref{fig:ari_sect_ncl} a) refers to the non-detrended case, Fig. \ref{fig:ari_sect_ncl} b) to the detrended case. The vertical dashed line in the graphs identifies the value $N_{cl} = 19$, that is the number of ICB supersectors.
For all the methods we observe an increasing trend for low values of $N_{cl}$, a maximum
and then a decreasing trend toward zero as $N_{cl}$ goes to 342. However the five methods show differences for what concerns the value of the maximum and its position. 
In the non-detrended case, we find that the highest values of Adjusted Rand Index, $\mathcal{R}^*_{adj}$, are reached by DBHT ($0.419$), k-medoids ($0.387$) and Complete Linkage ($0.387$). Interestingly these three values are quite close to each other,
maybe indicating this level as the 
actual maximum similiarity between correlation clustering and ICB supersectors. However the number of clusters correspondent to each maximum ($N^*_{cl}$) depends on the method and is respectively 13, 17 and 39. It is worth noticing that the maximum for the 
DBHT and k-medoids occurs very close to the ``real'' ICB supersectors $N_{cl}=19$ indicated by the dashed line in Fig.\ref{fig:ari_sect_ncl}. However, the k-medoids values are sensitively
 more fluctuating than the two hierachical methods. The Average Linkage and Single Linkage reach
instead much lower $\mathcal{R}^*_{adj}$ (respectively $0.352$ and $0.184$) and much higher $N^*_{cl}$ (respectively 111 and 229).  

For what concerns the detrended case, we notice first of all that the maximum values of $\mathcal{R}_{adj}$ increase for all the methods. The natural explanation for this is that
 the market mode, driving all the stocks regardless of their industrial supersector, hides to some extent the ICB structure \cite{market_mode}. The CL shows now the highest degree of similarity ($0.510$), followed by the AL
  ($0.48$, showing the most remarkable increase with respect to the non-detrended case), the k-medoids ($0.467$) and DBHT ($0.444$). The SL is again the last one in the ranking ($0.315$).
  The ranking in $N^*_{cl}$ is instead the same of the market mode case: lowest $N^*_{cl}$ for DBHT (20), followed by k-medoids (25), Complete Linkage (50), Average Linkage (60) and Single Linkage (101). Let us stress that, although the DBHT has not the highest $\mathcal{R}_{adj}$ in this case,
 its maximum is the closest to the real $N_{cl}$ of 19. In general the effect of the market mode subtraction on $N^*_{cl}$ changes according to the clustering method: for the DBHT, CL and k-medoids the subtraction raises $N^*_{cl}$,
  whereas for AL and SL the same quantity reduces remarkably. The effect is more pronunced for the Linkage methods, less for DBHT and k-medoids.
 
 Overall we can conclude that varying the number of clusters ($N_{cl}$) the DBHT, k-medoids and CL outperform the other two clustering methods at retrieving the ICB information. 
 DBHT, k-medoids and CL have however peaks of $\mathcal{R}^*_{adj}$  at different $N^*_{cl}$ values. 
 In particular, the DBHT and k-medoids are able to retrieve the ICB information at a $N^*_{cl}$ that is both lowest and closest to the actual number of ICB supersectors (19).
 After subtracting the market mode also the AL reaches the same level of DBHT, k-medoids and CL, but at too high $N^*_{cl}$.
 
 Interestingly, the Average and Single Linkage methods have the clusterings with both the lowest values of $\mathcal{R}_{adj}$ and the highest disparity values $y$:
 i.e. it appears that the higher the disparity $y$ is, the less the clustering method is able to retrieve the industrial classification. This has to be due to the presence of a large cluster when $y$ is very high: this again indicates a strong
 sensitivity to the market mode {\color{black} (see previous Section)}, that hides the intrasector correlations merging many stocks in a single cluster.

\subsubsection*{Industries overexpression}

\begin{figure}[ht!]
\begin{center}
\begin{tabular}{l}
  \includegraphics[scale=1.35]{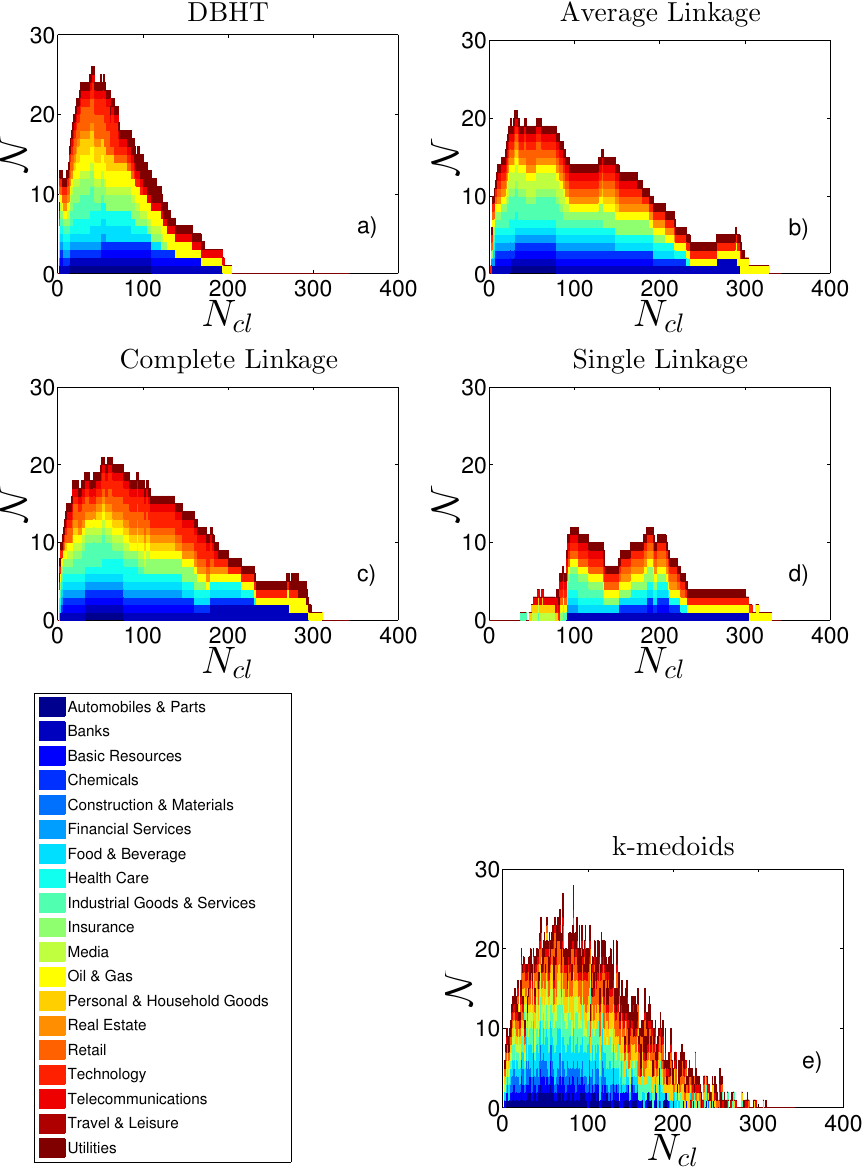}
\end{tabular}
\end{center}
 \caption{\label{fig:histo_sect_ncl} {\bf Amount of ICB information retrieved by the clustering methods, in terms of ICB supersectors overexpressed by each cluster}. Each bar graph shows, varying the number of clusters $N_{cl}$, how many times ($\mathcal{N}$) 
 an ICB {\color{black} supersector} is overexpressed by a cluster according to the Hypergeometric hypothesis test
  (i.e., number of null-hypothesis tests being rejected). 
  Each colour shows the number of overexpressions for each ICB {\color{black} supersector}. 
  In graphs a)-e) the results for DBHT, AL, CL, SL and k-medoids clustering are shown respectively.   
  The correlations are calculated on detrended log-returns. See Fig. S5 in S1 file for the non-detrended case.} 
  %See Fig. \ref{fig:histo_sect_ncl_subtracted} and Fig. \ref{fig:histo_sect_ncl_supersectors} in the \ref{appendix:hypergeometric_test} for the detrended cases and with ICB supersectors. 
\end{figure}

 {\color{black} As we have stated above, the Adjusted Rand Index provides an overall measure of similarity between the clustering and the ICB partition. Let us now focus on a more refined level of analysis, with which we quantify
 to what extent each industrial sector is retrieved by the clustering}. 

% In order to have a smaller number of ICB communities to compare with the clusters, let us consider here the ICB industries instead of the ICB supersectors. This means looking at the higher hierarchical level in the ICB classification,
% that gathers the stocks in 10 different industries. Let us remark that we have performed these analyses also on ICB supersectors, and the results are comparable (see \ref{appendix:hypergeometric_test} for the results).

 {\color{black} To this aim we have varied} the number of clusters $N_{cl}$ and, for each clustering, we have performed {\color{black}  a one-tail hypothesis test (see  Methods, Eq. \ref{eq:hypog_test})} for each pair cluster/ICB 
 supersector. 
 In Fig. \ref{fig:histo_sect_ncl} we show with bar graphs the results of these analyses
 for each one of the five clustering methods, by using detrended log-returns. 
 The x-axis shows $N_{cl}$, whereas $\mathcal{N}$ on the y-axis represents the total number of hypergeometric tests that have been rejected for that value of $N_{cl}$ (i.e. how many times an ICB {\color{black} supersector} has been found to be overexpressed by a cluster).
 The colors on each bar show the number of overexpressions for each different {\color{black} supersector}. We have chosen a significance level for the test equal to 0.01, together with the conservative Bonferroni correction \cite{feller2008introduction} that reduces the significance level to $0.01/(0.5*N_{cl}*N_{ICB})$, with $N_{ICB}$ being the number of ICB supersectors ($19$).
 (In Fig.S6 in S1 file, we report also the relative numbers  $\mathcal{N}/(0.5*N_{cl}*N_{ICB})$.)
 
 As we can see, the compositions and trends with $N_{cl}$ are quite different among different methods. The DBHT shows the highest number of clusters overexpressing ICB industries, followed by k-medoids, CL, AL and SL.
Let us note that AL and SL show the worst performances also in the Adjusted Rand Index analysis in Fig. \ref{fig:ari_sect_ncl} a). 
However, in that analysis the DBHT, k-medoids and CL were showing very close Adjusted Rand Index values;
 the hypothesis test is therefore able to highlight better the differences between these methods. For what concerns the industry composition, we can see that the DBHT, k-medoids, CL and AL show a quite homogeneous composition,
 with almost each ICB supersector overexpressed. The SL instead shows a much less rich composition, with no more than 6 overexpressed supersectors simultaneously even at the maximum level of total overexpressions. 
 
 In terms of $\mathcal{N}$ shape the DBHT is quite peaked, quickly dropping to low values for high $N_{cl}$ values. The three Linkage methods are instead flatter and more spread along the $N_{cl}$ axis, a feature 
 that was evident also in the trend of $\mathcal{R}_{adj}$ in Fig. \ref{fig:ari_sect_ncl}. The k-medoids seems to be a mix between these two shapes, showing however a much higher level of noise and instability in the ICB 
 composition when $N_{cl}$ changes.
 
 Finally, it is worth noticing that there is a change in the composition at different values of $N_{cl}$, and that similar patterns can be found among the four hierarchical clustering methods (for the k-medoids no clear patterns can
 be found, because of the higher level of instability of the method).
 There are {\color{black} supersectors} that tend to become overexpressed for low values of $N_{cl}$ and then disappear at intermediate values: this is the case of {\color{black} Automobiles \& Parts, Telecommunications, Insurance and Financial Services}.
 Others are instead more persistent, appearing 
 along all the x-axis: Utilities, Technology, Oil \& Gas. The most persistent is the latter, that is still overexpressed when all the other {\color{black} supersectors} are not expressed anymore. We can then conclude that not only the ICB 
 partition is hidden at different levels in the dendrograms (see ``Retrieving the industrial sectors'' Section) depending on the clustering method, but also different ICB {\color{black} supersectors} are retrieved at different levels. 
 This is probably due to the different degrees of correlation within different ICB {\color{black} supersectors}.  
 
 By using {\color{black} non-detrended} log-returns we obtain quite similar results, apart from an overall {\color{black} decrease} in $\mathcal{N}$ for all the methods, consistently with what found in ``Retrieving the industrial sectors''. 
 The details are shown
 in Fig. S5 in S1 file.
 %in \ref{appendix:hypergeometric_test}, Fig. \ref{fig:histo_sect_ncl_subtracted}.

\subsection*{Dynamical analysis}
\label{sec:dyn_analysis}

\begin{figure}[ht!]
\begin{center}
\begin{tabular}{l}
  \includegraphics[scale=0.95]{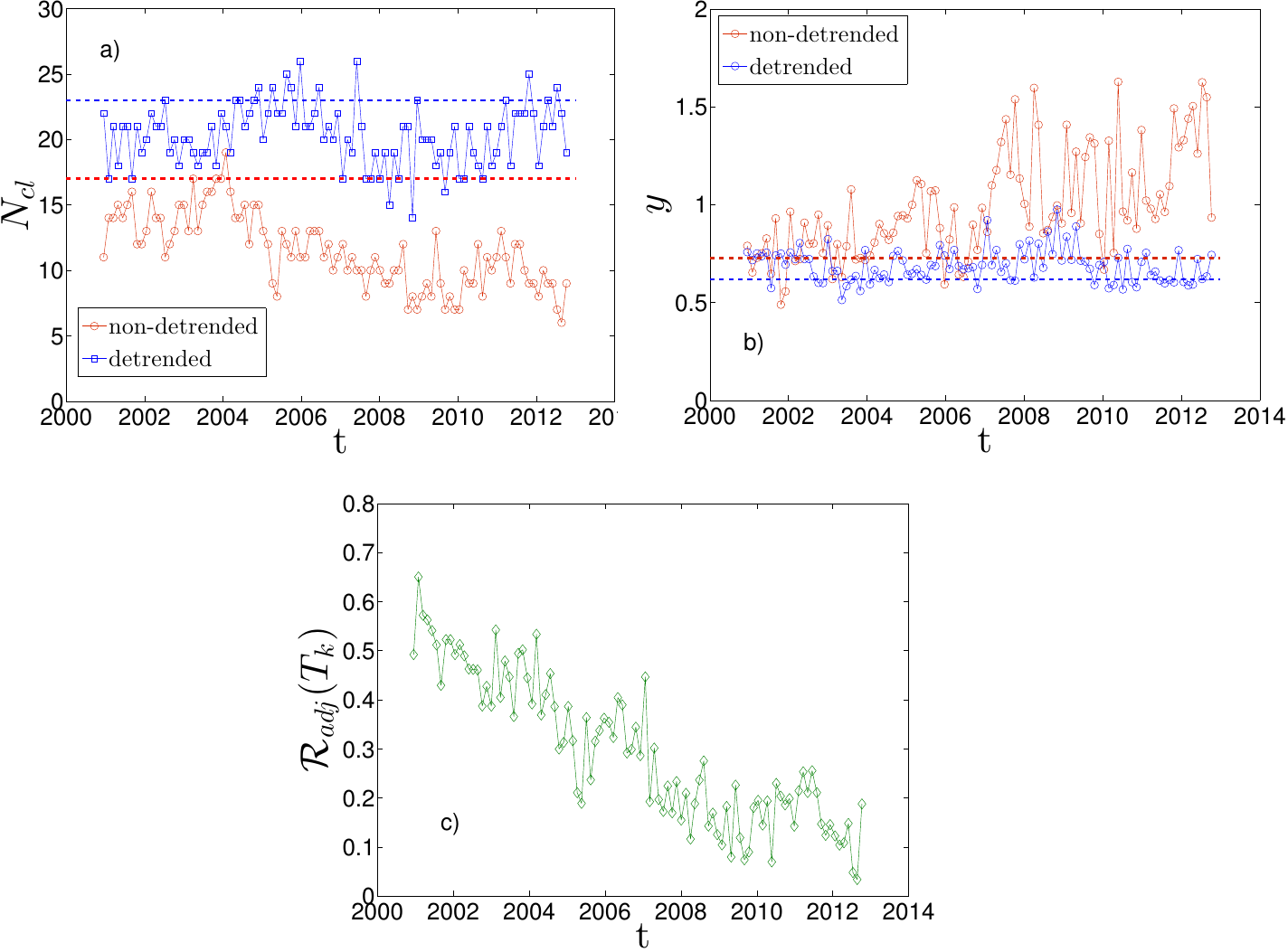}
\end{tabular}
\end{center}
 \caption{\label{fig:num_cl_dbht} {\bf Dynamical evolution of the DBHT clustering}. Each plot refers to 100 moving time windows of length 1000 trading days.
 Specifically, in graph a) we plot the number of DBHT clusters, $N_{cl}$, for both log-returns non-detrended (red circles) and detrended by the market mode (blue squares), whereas the two dashed horizontal lines are the $N_{cl}$ values 
 obtained by taking the largest time window of 4026 trading days. Overall the non-detrended case shows a decreasing trend. In graph b) it is shown the disparity measures, $y$, again for the two sets of DBHT clustering (red dots non-detrended, blue dots detrended), the dashed horizontal
 lines being the $y$ values from the 4026 length time window. In the non-detrended case the 2007 marks a transition to higher and more volatile values of $y$. Finally in graph c) it is shown the Adjusted Rand Index,
 $\mathcal{R}_{adj}$, measured at each time window between the detrended and non-detrended clusterings. A steady decreasing trend is evident.} 
\end{figure} 

Here we present a dynamical analysis of the DBHT clustering in the 15 years period ranging from 1 January 1997 to 31 December 2012. We have selected the set of overlapping time windows described in ``Dataset and preliminary analyses''
($n=100$ time windows of length $L=1000$ trading days) and used a weighted version of the Pearson estimator (Eq. \ref{eq:correlation}) in order to mitigate
 excessive sensitiveness to outliers in remote observations. 
 
 In Fig. \ref{fig:num_cl_dbht} a) is shown 
the number of DBHT clusters obtained for each time window, both for non-detrended log-returns (red circles) and detrended log-returns (blue squares). For the first case the number of clusters ranges between 6 and 19, for the second 
case the range is 14-26. {\color{black} Let us stress that this is the ``natural'' number of clusters automatically provided by the DBHT method, and it is not the result of any thresholding}. 
Dashed lines are the values correspondent to the clustering obtained using the entire time period 1997-2012 as time window. 
A study of the statistical robustness of these clusterings have been carried out by means of a bootstrapping approach: it turns out that the number of clusters shown in Fig. \ref{fig:num_cl_dbht} a) 
is robust against resampling of the log-returns time series, as reported in the next section ``Robustness of DBHT: a bootstrapping analysis''.

As observed previously, the number of clusters in the non-detrended case  
is sistematically lower than in the detrended case. Moreover, an overall decreasing trend characterizes the non-detrended values and makes them go below the corresponding dashed line; this decreasing pattern is not present in 
the detrended case, that however stays below the correspondent dashed line most of the times either.  
The evolution of the disparity $y$, introduced in Eq. \ref{eq:disparity}, is shown in Fig. \ref{fig:num_cl_dbht} b) both for non-detrended and detrended case. 
Again the dashed lines are the values for the all period. 
In the non-detrended case we see an overall increasing trend, especially after the 2006; an analysis of the sizes distribution 
shows that largest cluster contains up to 240 stocks ($70\%$ of total number of stocks); moreover, from 2006 onwards we observe also a much higher fluctuation in the values. 
This behaviour is of interest since it concerns the overall influence of the market mode on the correlation 
structure, with higher $y$ indicating a stronger influence of the market mode that tends to gather all stocks in one cluster. Indeed, in the detrended case we find that subtracting
the market mode makes the increasing trend disappear with decreasing disparity values that stay closer to the dashed line, without significant patterns apart from some fluctuations.

In order to better understand the relation between the DBHT clusterings obtained with detrended and non-detrended log-returns, we have also performed a dynamical Adjusted Rand Index analysis. Now we compare no longer the clustering 
and the ICB partition, but the two clusterings (non-detrended and detrended) at each time window. In Fig. \ref{fig:num_cl_dbht} c) the Adjusted Rand Index between the two sets of DBHT clusters is shown.
Interestingly, it shows a steady decreasing trend that drives the similarity from relatively high values (about $0.7$) to values close to zero, indicating complete uncorrelation between the two clusterings. We can therefore conclude that the influence of the market mode has increased remarkably over the last $15$ years, making the detrended clustering structure more and more different from the non-detrended one. 
Let us note that this observation would have not been possible without the clustering analysis, since from the preliminary dataset measures (see Fig. \ref{fig:avg_corr_vol} and Fig. S2 in S1 file for details)  it is not evident any constant pattern either in the average return or in the average correlation.

\subsubsection*{Dynamically retrieving the industrial sectors }
\label{subsec:dyn_ind_sectors}

Let us here investigate the relation between industrial classification and clustering under a dynamic perspective.
 To this end we here perform the previous dynamical analysis by considering the set of 100 overlapping time windows $T_k$ 
 and calculating for each of them the Adjusted Rand Index $\mathcal{R}_{adj}(T_k)$ between clustering and ICB supersectors classification. Since $\mathcal{R}_{adj}(T_k)$ varies with the chosen threshold and $N_{cl}$, we select at every time the $N_{cl}$ that maximizes $\mathcal{R}_{adj}(T_k)$; the numbers that we report are these maximum values and account therefore for the maximum ability of the clustering methods to retrieve the ICB. 
 
\begin{figure}[ht!]
 \begin{center}
\begin{tabular}{l}
  \includegraphics[scale=0.92]{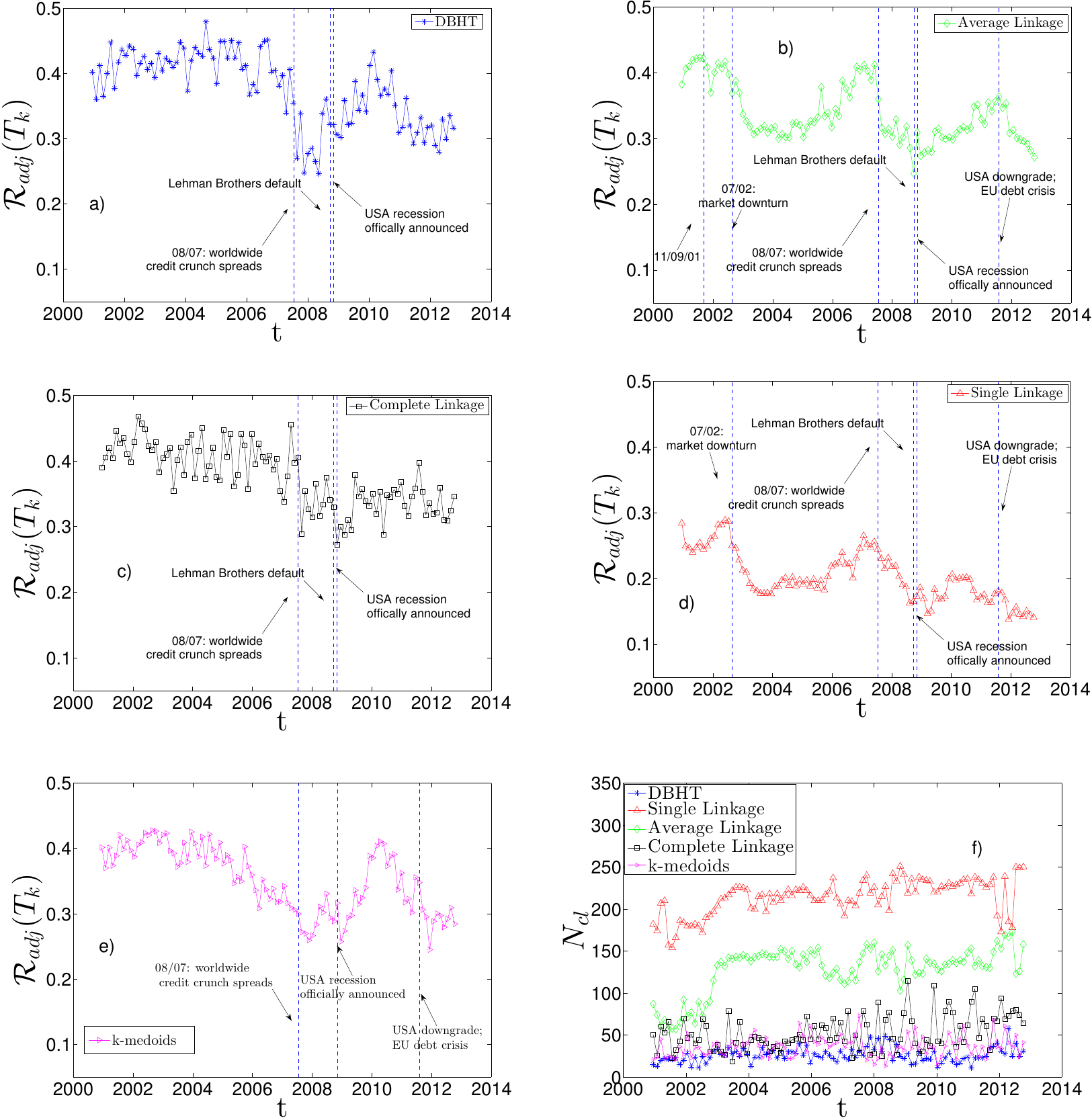}
\end{tabular}
\end{center}
 \caption{\label{fig:ari_sect} {\bf Dynamical evolution of the similarity between clustering and ICB}. It is shown the Adjusted Rand Index, $\mathcal{R}_{adj}$, calculated at each time window $T_k$ ($k=1,...,n$) between clustering and ICB partition,
  for a) DBHT, b) AL, c) CL, d) SL and e) k-medoids method. A drop in the similarity occurs for all the methods during the 2007-2008 crisis. 
 The AL and SL show decreases also during other financial events.    
  At each time window the number of clusters, $N_{cl}$, has been chosen in order to maximize the $\mathcal{R}_{adj}$ itself:
  in f) we plot these $N_{cl}$ values for each clustering method. It is evident as the maximum similarity clustering-ICB is reached at different hierarchical levels depending on the clustering method.
   The correlations are calculated on non-detrended log-returns.} 
\end{figure}

 In Figs. \ref{fig:ari_sect} a)-e) we show the results for each of the five clustering methods, using returns with market mode. Interestingly, all of them show a decreasing trend across the time. On average, the DBHT and CL
 display the highest similarity with industrial classfication, whereas the Single Linkage the lowest. This is consistent with what found in the static analyses.
   We have also highlighted in the graphs the major events that affected the stock market in the last 15 years. It can be observed that different clustering methods are affected in different ways by these events. 
   For instance we observe that, if the  2007-2008 crisis and the following recession is evident in all the methods as a significant drop in the similarity, other events such as 11/09/2001 or the 2002 stock market downturn appear only in the Single and Average Linkage plots.
In particular the 2002 downturn drives a steep decrease in the similarity of SL and AL, staying at low values until the end of 2005. Instead DBHT, Complete Linkage and k-medoids do not seem to be significantly affected by
these events. 
    This observation points out that the DBHT, CL and k-medoids are more robust than SL and AL against exogenous events in their ability to retrieve an economic information as the industrial classification. 
    Nonetheless, there are differences also among DBHT, CL and k-medoids: in particular in the period following the 2008 crisis, DBHT and k-medoids show a peak that does not appear with CL. Moreover, for the k-medoids 
    the drop in similarity seems to start more than one year before the 2007. All these features have non-trivial implications for both portfolio optimization and systemic risk evaluation. We plan to investigate these implications in 
    future works.
  
  Fig. \ref{fig:ari_sect} f) shows the number of clusters $N_{cl}$ that, in each time window $T_k$, maximizes the Adjusted Rand Index shown in the previous plots. As we can see, $N_{cl}$ for SL is always the highest, followed by AL, CL, k-medoids and DBHT. 
This is consistent with what we found in the ``Static analysis'' Section: different clustering methods ``hide'' the industrial information at different levels of the hierarchy. 
SL and AL, that yield higher $N_{cl}$ (i.e., lower levels in the hierarchy), are also the methods that show the lowest level of similarity with industrial classification and the highest degree of disparity.
   
\begin{figure}[ht!]
 \begin{center}
\begin{tabular}{l}
  \includegraphics[scale=0.92]{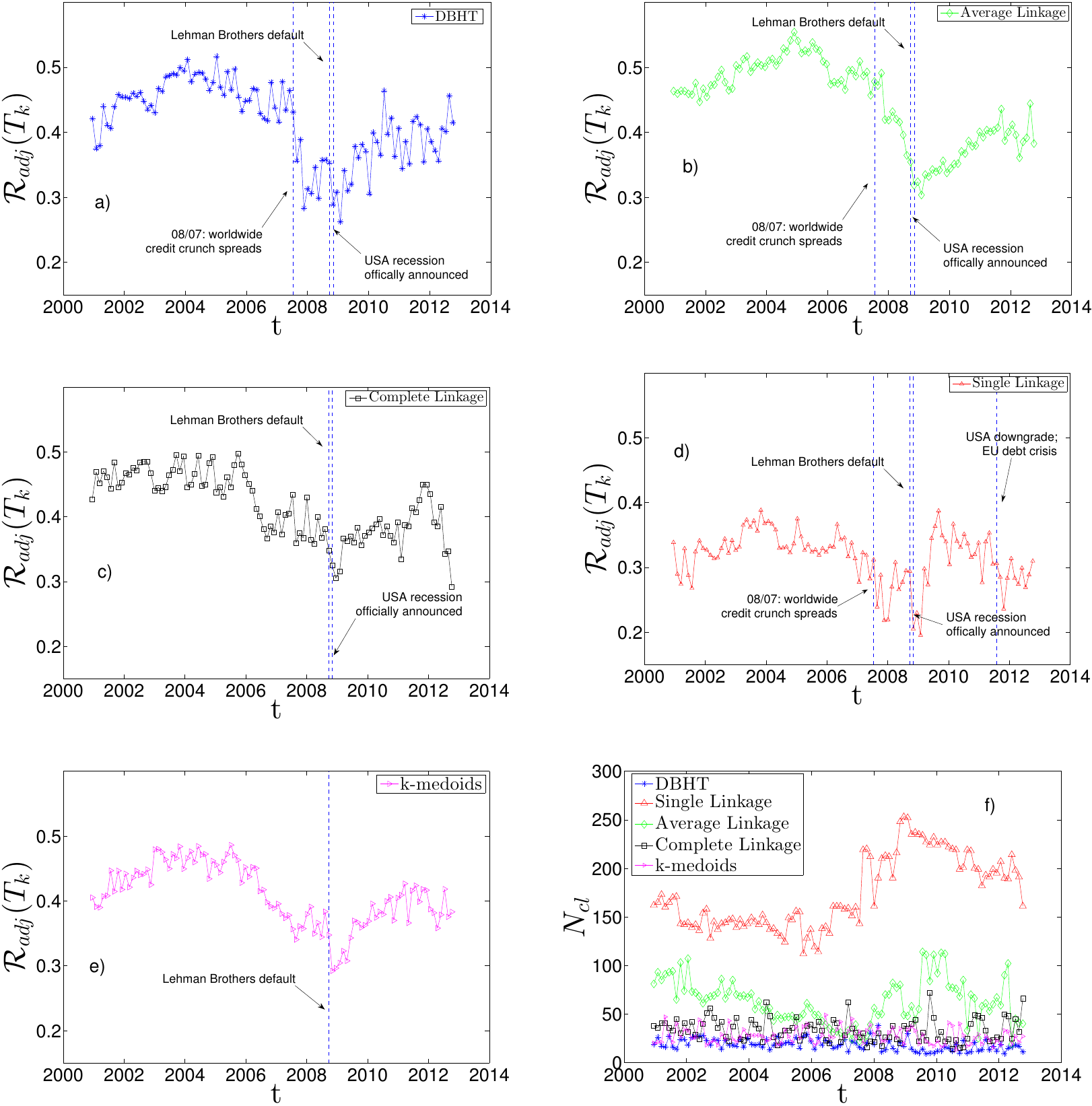}
\end{tabular}
\end{center}
 \caption{\label{fig:ari_sect_subtracted} {\bf Dynamical evolution of the similarity between clustering and ICB, with detrended log-returns}.
 a)-f): Same graphs as in Fig. \ref{fig:ari_sect}, but by using correlations on detrended log-returns.} 
\end{figure}

   In Figs. \ref{fig:ari_sect_subtracted} a)-f) we show the same set of plots for the detrended case. The main differences with the non-detrended case are the following:
   
   \begin{itemize}
    \item the average similarity with the industrial classification rises for all methods; this confirms in the dynamical case what we found for the static case;
    
    \item the average $N_{cl}$ is lower for all methods: the absence of market mode ``moves'' the industrial classification to higher levels of the hierarchy;
    
    \item the strong influence of the 11/09/2001 and 2002 downturn on the SL and AL pattern seems to disappear, whereas the 2007-2008 crisis is still evident in all the five methods. This could be explained claiming that the formers
          are global events in the market, whereas the latter exhibits also a ``local'' dynamics;
          
    \item the AL shows the most evident change in the dynamical behaviour, displaying a trend much more similar to the DBHT and CL one. Also in terms of $N_{cl}$, it shows values closer to DBHT, CL and k-medoids than SL. A similar observation
          was made in the static analysis in ``Retrieving the industrial sectors'', where the AL turned out to perform like DBHT, CL and k-medoids once the market mode was removed.
          
   \end{itemize}

\subsubsection*{Robustness of DBHT: a bootstrapping analysis}

\begin{figure}[ht]
 \begin{center}
\begin{tabular}{l}
  \includegraphics[scale=1]{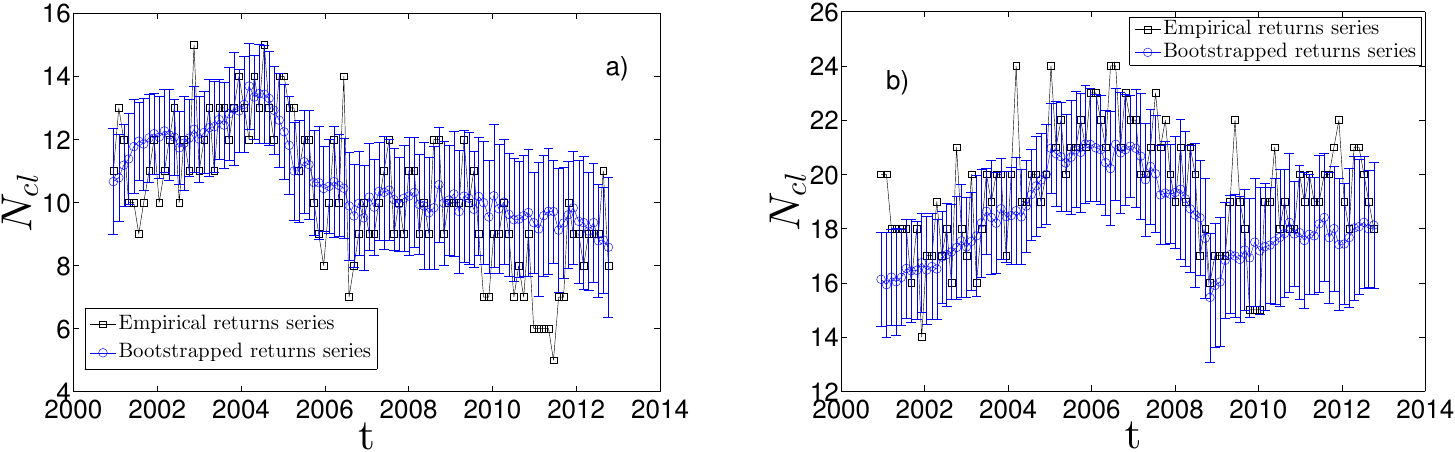}
\end{tabular}
\end{center}
 \caption{\label{fig:bootstrapping} {\bf Test of robustness for the dynamical DBHT clustering}. a) Number of clusters $N_{cl}$ as a function of the time $t$: the black squares correspond to the DBHT clusterings obtained by using the empirical (non-detrended) log-returns, the blue dots are the average over the 100 $N_{cl}$
  given by the 100 bootstrapping replica correlation matrices (see text for further details). The bar errors in the blue dot plot is the standard deviation calculated among the same set of 100 $N_{cl}$. As one can see the empirical $N_{cl}$
   is quite robust against the bootstrapping test. b) Same plot as in a), but by using detrended log-returns.} 
\end{figure}

In order to test the sensitiveness of the DBHT clustering to the statistical noise, inevitably present in every correlation estimate, we have performed the Bootstrapping test to our dataset \cite{bootstrapping}. 
 {\color{black} This method generates a set of $n_{boot}$ replicas of each original correlation matrix by randomly resampling the log-returns matrix. On each replica of correlation matrix we can calculate a new DBHT clustering, ending up
 with a sample of $n_{boot}$ clusterings on which we can perform statistical analyses.
A detailed description of the Bootstrapping method can be found in SI}.

In Fig. \ref{fig:bootstrapping} we show the result of the dynamical Bootstrapping performed over all the 100 time windows that cover the entire period. We chose $n_{boot}=100$: each time window has therefore $100$ replica DBHT 
clustering associated.
The blue points are the average number of clusters over the $n_{boot}$ replicas of DBHT clusterings, whereas the error bars are the standard deviations calculated over the same sample.
The black squares are the empirical numbers of clusters 
yielded by the DBHT. The left-hand side plot (a)) is obtained by using non-detrended log-returns, the right-hand side (b)) is obtained by using detrended log-returns.

The plot of empirical number of clusters is slightly different from what we have shown in Fig. \ref{fig:num_cl_dbht} a) because for this bootstrapping analysis we did not use exponential smoothing for the correlations, but only 
bare correlations. The exponential smoothing, indeed, creates an asymmetry among the points in each time series that makes the bootstrapping test inapplicable. {\color{black} Overall, removing the exponential smoothing affects each
$N_{cl}$ value
 by $20\%$ on average}.

From the plot we can observe that the method is statistically robust, with the most of empirical points within one standard deviation from the mean of replicas. More importantly, the mean of replicas follows the general trend 
of the empirical points; namely, the decreasing trend in the market mode case, and the drop after the 2007-2008 {\color{black} crisis} in the detrended case.

\subsection*{Discussion}
\label{conclusion}
In this paper we have presented a set of static and dynamical analyses to quantify empirically the amount of information filtered from correlation matrices by different hierarchical clustering methods.
By taking the Industrial Classification Benchmark (ICB) as a benchmark community partition we have performed quantitative analyses on real data without any assumption on the returns distribution.

 In particular we have considered three variants of Linkage methods (Single, Average and Complete) and the k-medoids, and we have compared them with the Directed Bubble Hierarchical Tree (DBHT), a novel clustering
 method applied here for the first time to financial data. 
%   Our analyses have shown that the DBHT is a more suitable 
%  clustering method than the Linkage and the k-medoids for financial data, being able to retrieve more information with fewer clusters.
 
 The methods show remarkably different performances in retrieving 
 the economic information 
 encoded in the ICB, with big dissimilarities even among the Linkage methods. {\color{black} We have suggested that these differences should be connected to different degrees of sensitivity to the market mode dynamics, that in turns 
 are to be ascribed to differences in the methods underlying working principles}. Moreover, the economic information appears to be hidden at different levels of the hierarchical structures depending on the clustering method. The DBHT 
 and k-medoids methods show the best performances, but the latter seems to be affected by the noise much more than the DBHT and the Linkage methods. The DBHT turns out then to be a good mix between 
 the advantages of the k-medoids and those of the Linkages. The dynamical analysis has also proved that the methods show different degrees of sensitivity to financial crises. This is again a new result
 that could give insights into the dynamics of such events, as well as an indication on which clustering method is more robust for financial applications.
 
 We have also performed each analysis on log-returns detrended by the market mode, by following a standard procedure in literature \cite{market_mode} \cite{multifactor_clustering}.
 Interestingly the effect of this {\color{black} detrending} is very dissimilar 
 for different methods, with the weakest methods (Average and Single Linkage) improving remarkably their {\color{black} ability to retrieve industrial sectors}. {\color{black} In general the detrending increases the degree of economic information 
 that the clustering methods retrieve. It also makes the clusterings more homogeneous in sizes, suggesting that the high heterogeneity in SL and AL must be due to the market mode dynamics.}
 %However the main results are robust against the subtraction of the market mode.  
 
 In future works we plan to extend the present study to other datasets, covering different periods and different stock exchanges and considering other measures of dependences including non-linear dependences such as
 the Kendall's rank correlation \cite{kendall} and the Mutual information \cite{mutual_info} \cite{fiedor_mutual_info}.
   Finally, since correlation-based networks and clustering methods have shown to be useful tools for portfolio optimization \cite{invest_periph},
   \cite{clustering_indian_mkt,kmeans_portfolio}, we also plan to use these new 
 insights into the hierarchical structures to improve further the current performances of portfolio optimization tools.

\section*{Acknowledgments}
The authors wish to thank Bloomberg for providing the data.
TDM wishes to thank the COST Action TD1210 for partially supporting this work. TA acknowledges support of the UK Economic and Social Research Council (ESRC) in funding the Systemic Risk Centre (ES/K002309/1).

% The bibtex filename
%\bibliographystyle{plain}
\bibliographystyle{plainnat}
\bibliography{bibl_sample.bib}

\section*{Supporting Information}

\subsection*{S1. Dataset analysis}
\label{appendix:dataset}

 \begin{figure}[h!]
\begin{center}
\begin{tabular}{llc}
 \includegraphics[scale=0.4]{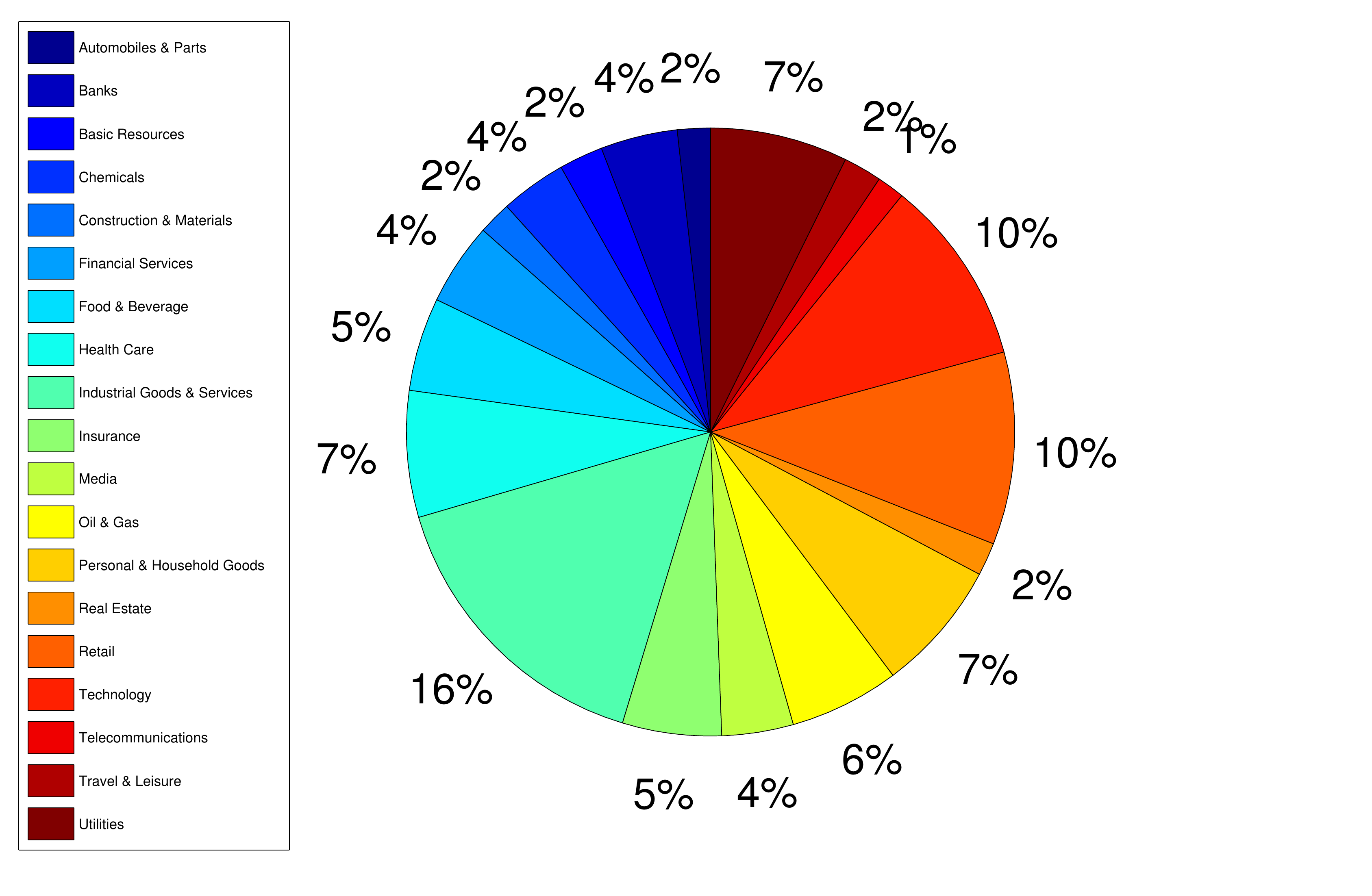} 
\end{tabular}
\end{center}
 \caption{\label{fig:pie_sectors} {\bf Pie chart showing the composition of the entire set of stocks in terms of ICB supersectors.} } 
\end{figure}
 
 The set of stocks has been chosen in order to provide a significant sample of the different
 industrial sectors in the market. We have chosen the ICB industrial classification, that yields 19 different Supersectors, that in turns gather in 10 Industries: the percentage of stocks belonging to each ICB supersectors is 
 reported in Fig. \ref{fig:pie_sectors} .
 
\begin{figure}[ht!]
\begin{center}
\begin{tabular}{llc}
\hspace{-3em}
 \includegraphics[scale=0.25]{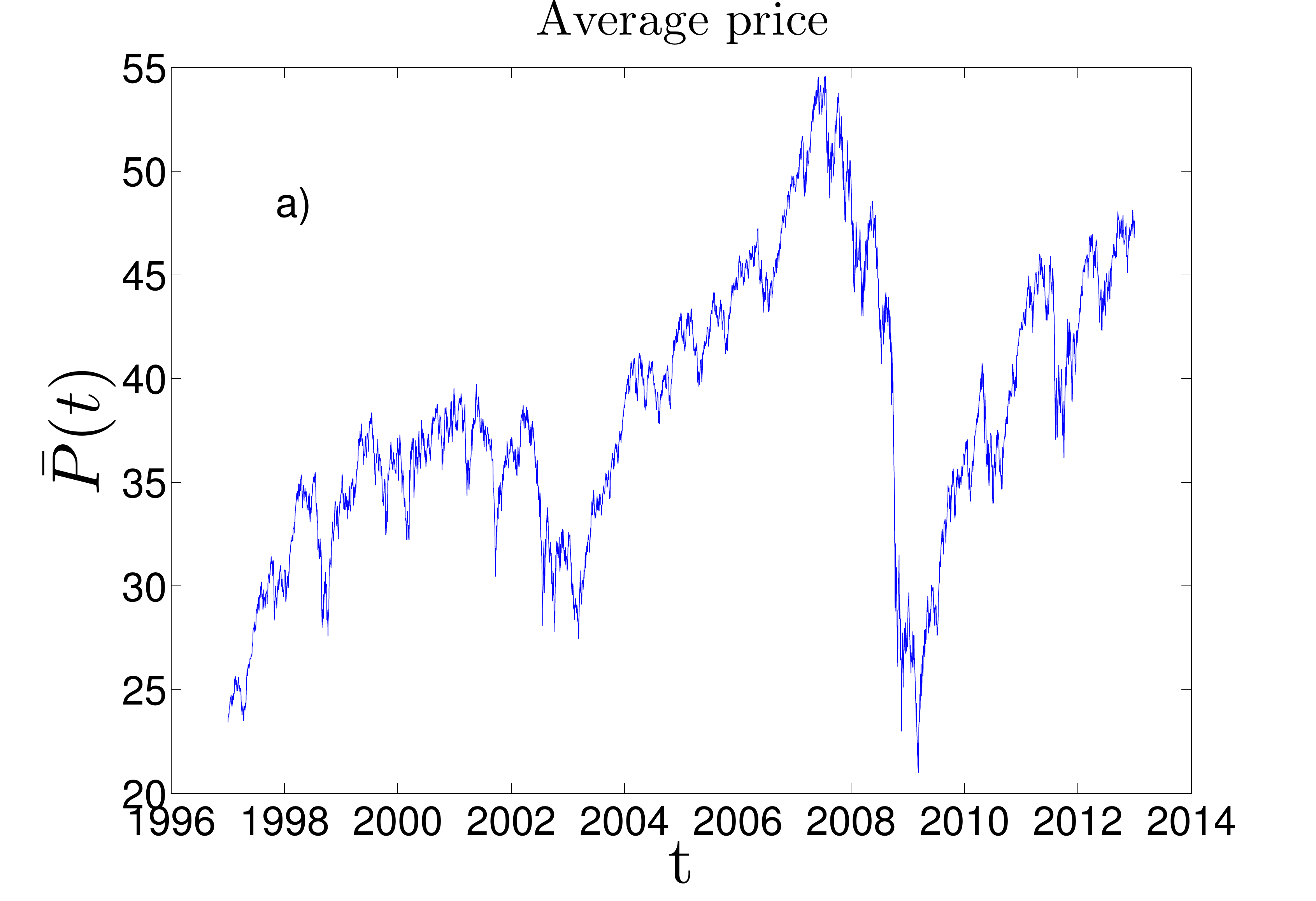} 
 \includegraphics[scale=0.25]{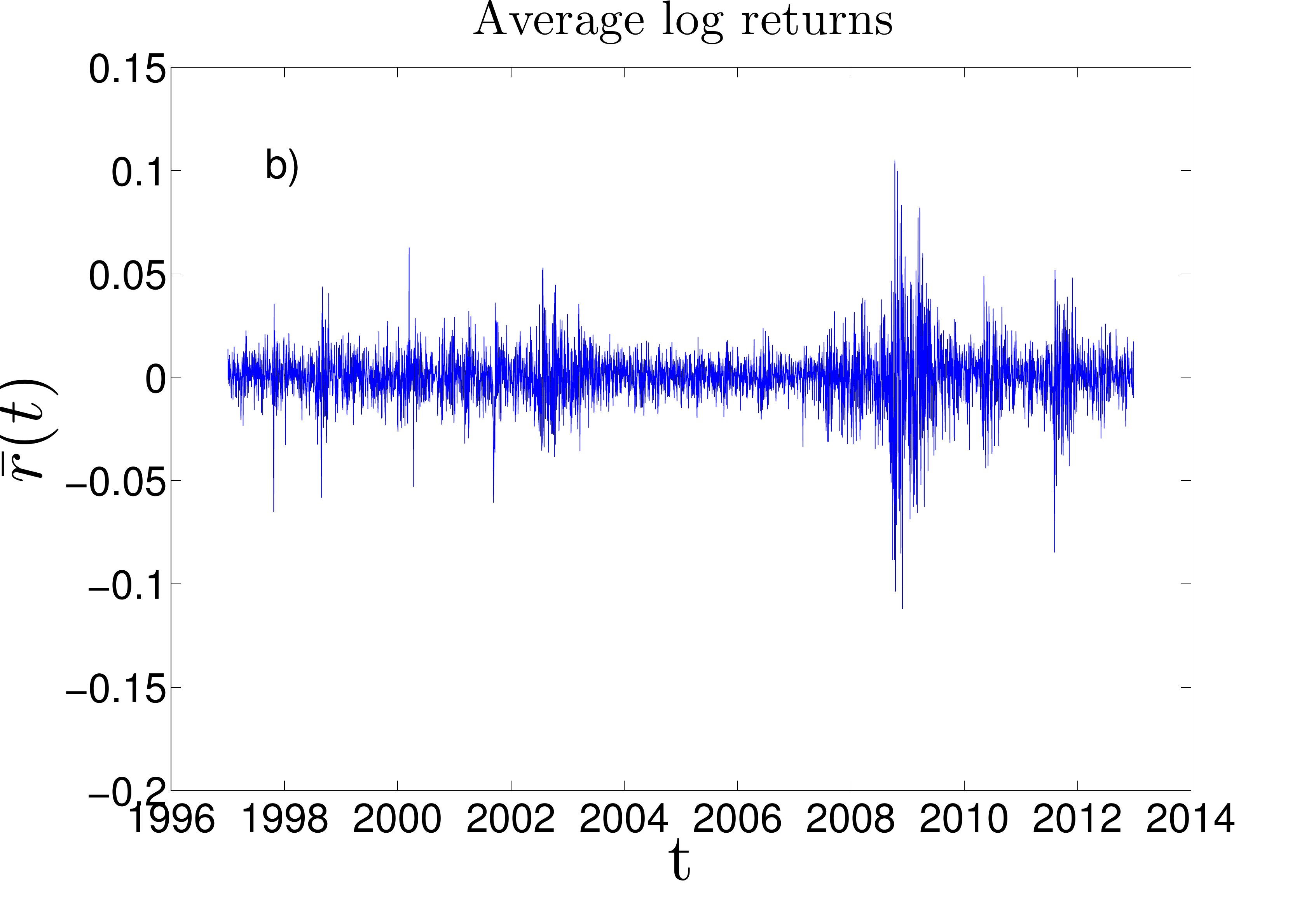}\\
\end{tabular}
\end{center}
 \caption{\label{fig:avg_prices} {\bf Average price and log-returns of the dataset, from January 1997 to December 2012}. a) Average price $\bar{P}(t)$ of the 342 US stocks in the dataset; b) Average log-return  $\bar{r}(t)$ 
 of the same prices.} 
\end{figure}

 In Fig. \ref{fig:avg_prices} two plots are shown that summarize the main features of this dataset. The graphs show the average price $\bar{P}(t) \equiv \frac{1}{N} \sum_iP_i(t)$ and 
 the average log return of the prices,
 $\bar{r}(t) \equiv \frac{1}{N} \sum_ir_i(t)$, as a function of time.
 From these plots we can see that both the internet bubble bursting (2002) and the credit crunch (2007-08) are displayed by the market dynamics. In particular it is evident a steep increase in volatility for both periods, 
 strongly autocorrelated in time: a well known feature of log-returns dynamics \cite{stylized_facts}. Such clusters of volatility can be observed also after the credit crunch, in 2010 and 2012.

\subsection*{S2. Clustering methods: a brief review}
 Our focus is on the hierarchical structures (dendrograms) that can best describe the interdependencies in the market, together with a cluster characterization of this structure. To this purpose, the first step is to define 
 a suitable distance between pairs of stocks, knowing the correlation between them. An appropriate function \cite{mantegna1} is $D_{ij}=\sqrt{2(1-C_{ij})}$ : it can be shown that with this choice $D_{ij}$ satisfies the three 
 properties of a distance measure.  
 
 We then end up with a set of $N \times N $ distance matrices $D(t_k)$ and $D^R(t_k)$, to which we apply two different, well known tools in order to reveal the hidden (unknown) structure of dependencies:
 
 \begin{itemize}
  \item \textbf{Single Linkage} (SL), that is an hierarchical clustering algorithm. Given the distance matrix, it starts assigning to each objects its own cluster, and then at each step merges the closest (i.e. least distant) 
  pairs of clusters into one new cluster, until only one cluster remains. The distances among two generic clusters $A$ and $B$ is everytime defined and updated according to the formula 
  
  \begin{equation}
  \label{eq:SL}
   d_{A,B}=\min_{a \in A, b \in B} D(a,b) 
  \end{equation}

  SL is called an \emph{agglomerative} clustering, since it begins with a partition of $N$ clusters and then proceed merging them. The final output of the method is a dendrogram, that is a tree showing the hierarchical structure
   found by the SL. The distance measure defined in this dendrogram is an ultrametric distance \cite{mantegna1}. 
   A proper cluster partition of the stocks can be obtained by choosing the number of clusters (that is therefore a free parameter) and cutting the dendrogram at the appropriate level. 
   
   This algorithm is strictly related to the one that provides a Minimum Spanning Tree (MST) given 
  the distance matrix $D$. The MST is a tree graph having the stocks as nodes, and it has been used as topological tool in Econophysics since the work of Mantegna \cite{mantegna1}. It can be generated starting with an empty graph: 
  after sorting all the correlations in $C$ in descending order, add a weighted link between the two stocks/nodes with the highest correlation, and then go ahead with the next highest pair correlation; whenever the new link to add 
  generates a loop, do not add that link and skip to the next one, until all the list is checked. 
  
  This tree contains exactly $N-1$ links.  It can be shown \cite{mantegna2} that the MST algorithm is basically the SL procedure carried out until the graph is completely connected. 
  There is therefore a strict relation between the two tools. However the MST retains some 
  information that the SL dendrogram throws away \cite{mantegna2} .
  
%   (MST) \cite{mantegna1}, that is a tree $MST(t_k)$ ($MST^{R}(t_k)$) associated to each given distance matrix $D(t_k)$ ($D^R(t_k)$). Each tree contains exactly $N-1$ links.
%   To each MST is naturally associated a dendrogram, generated by the linkage algorithm that runs in parallel with the MST construction . 

  \item \textbf{Average Linkage} (AL) is a hierarchical clustering algorithm similar to SL. The algorithm is the same as the one underlying the SL, but with Eq. \ref{eq:SL} replaced by:
  
   \begin{equation}
  \label{eq:SL}
   d_{A,B}= mean_{a \in A, b \in B} D(a,b) 
  \end{equation}
  
  \item \textbf{Complete Linkage} (CL) is a third variant of SL, where Eq. \ref{eq:SL} is replaced by:
  
    \begin{equation}
  \label{eq:SL}
   d_{A,B}=\max_{a \in A, b \in B} D(a,b) 
  \end{equation}
  
  \item \textbf{Directed Bubble Hierarchical Tree} (DBHT) \cite{DBHT}, a novel hierarchical clustering method that exploits the topological property of the PMFG (Planar Maximally Filtered Graph)  in order
  to find the clustering.
  
  The PMFG is a generalization of the MST, that is included in the PMFG as a subgraph. It is constructed following the same procedure of the MST, except that the non loop condition is replaced with 
  the weaker condition of planarity (i.e. each added link must not cut a pre-existent link). Thanks to this more relaxed topological constraint the PMFG is able to retain a larger amount of link, and then information, than the MST.
  In particular it can be shown that each PMFG contains exactly $3(N-2)$ links.
  
  The basic elements of a PMFG are three-cliques, subgraphs made of three nodes all reciprocally connected (i.e., triangles). The DBHT exploits this topological structure, and in particular the distinction between separating
  and non-separating three-cliques, to identify a clustering partition of all the nodes in the PMFG \cite{DBHT} . A complete hierarchical structure (dendrogram) is then obtained both inter-clusters and intra-clusters by 
  following a traditional agglomerative clustering procedure. 
  
  The Linkage algorithms look at the sorted list of distances $d_{ij}$ and build the dendrogram by gathering subsets of stocks with lowest distances; the clustering is then obtained, as we said, from the dendrogram 
  after choosing the parameter ``number of 
  clusters''. The DBHT instead reverses this order: first of all the clusters are identified by means of topological considerations on the planar graph, then the hierarchy is constructed both inter-clusters and intra-clusters.  
  The difference involves therefore both the kind of information exploited and the methodological approach.
    
  \item \textbf{k-medoids} is a partitioning clustering method closely related to k-means \cite{kmeans}. It takes the number of clusters $N_{cl}$ as an input. The algorithm is the so called Partitioning Around Medoids (PAM), and is as follows:
  
        \begin{enumerate}
         \item select randomly $N_{cl}$ ``medoids'' among the $N$ elements;
         \item assign each element to the closest medoid;
         \item for each medoid, replace the medoid with each point assigned to it and calculate the cost of each configuration;
         \item choose the configuration with the lowest cost;
         \item repeat 2)-4) until no change occurs.
        \end{enumerate}
 
  This method, alike the others taken into account here, is not a hierarchical method and does not provide therefore a dendrogram but only a partition.
   
 \end{itemize}

\subsection*{S3. Clustering compositions: non-detrended case}
\label{appendix:histo_cl}
  
\subsubsection*{DBHT clusters composition}
\label{appendix:histo_cl}

\begin{figure}[ht!]
\begin{center}
\begin{tabular}{l}
  \hspace{-5em}
  \includegraphics[scale=0.35]{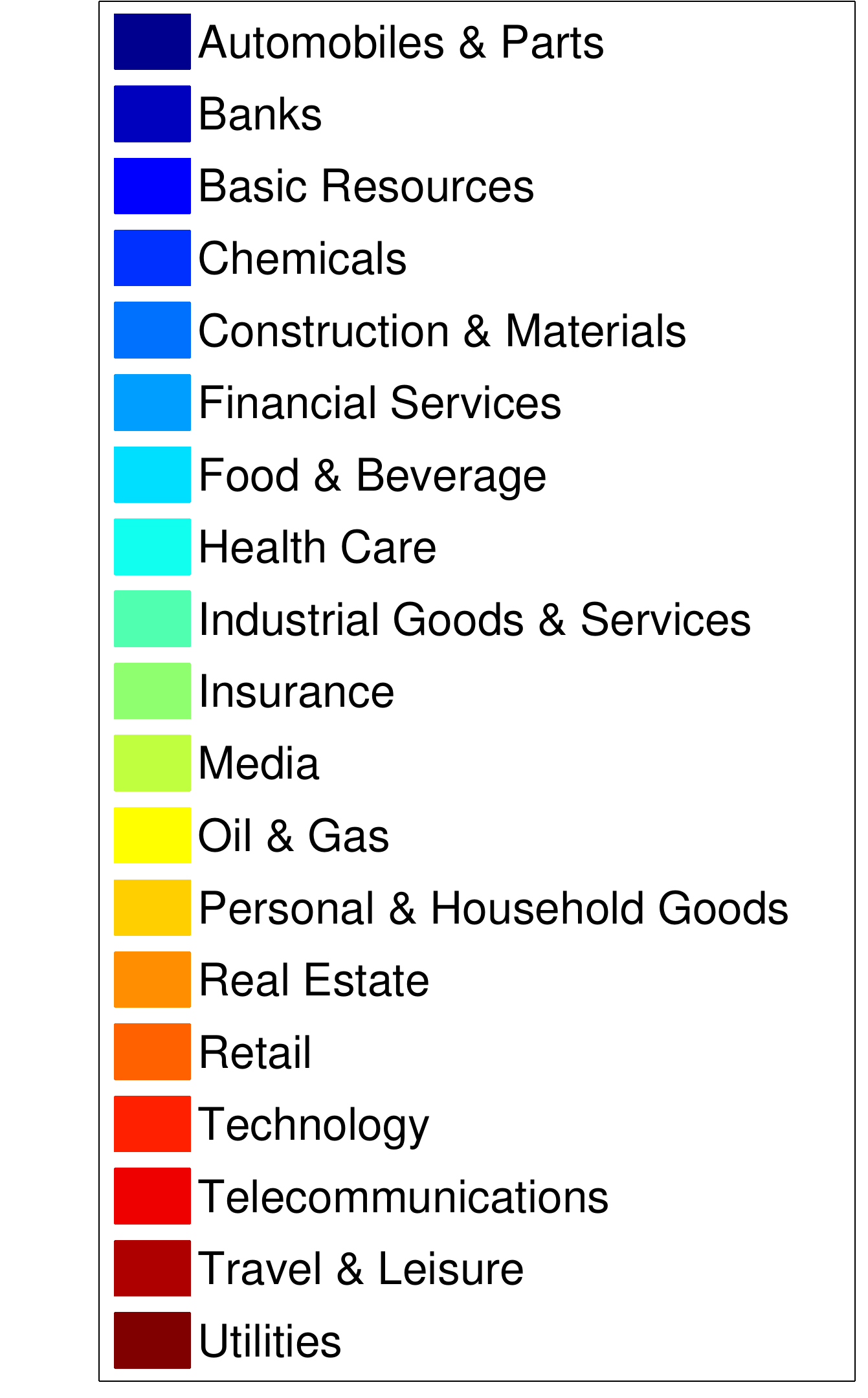}
  \includegraphics[scale=0.4]{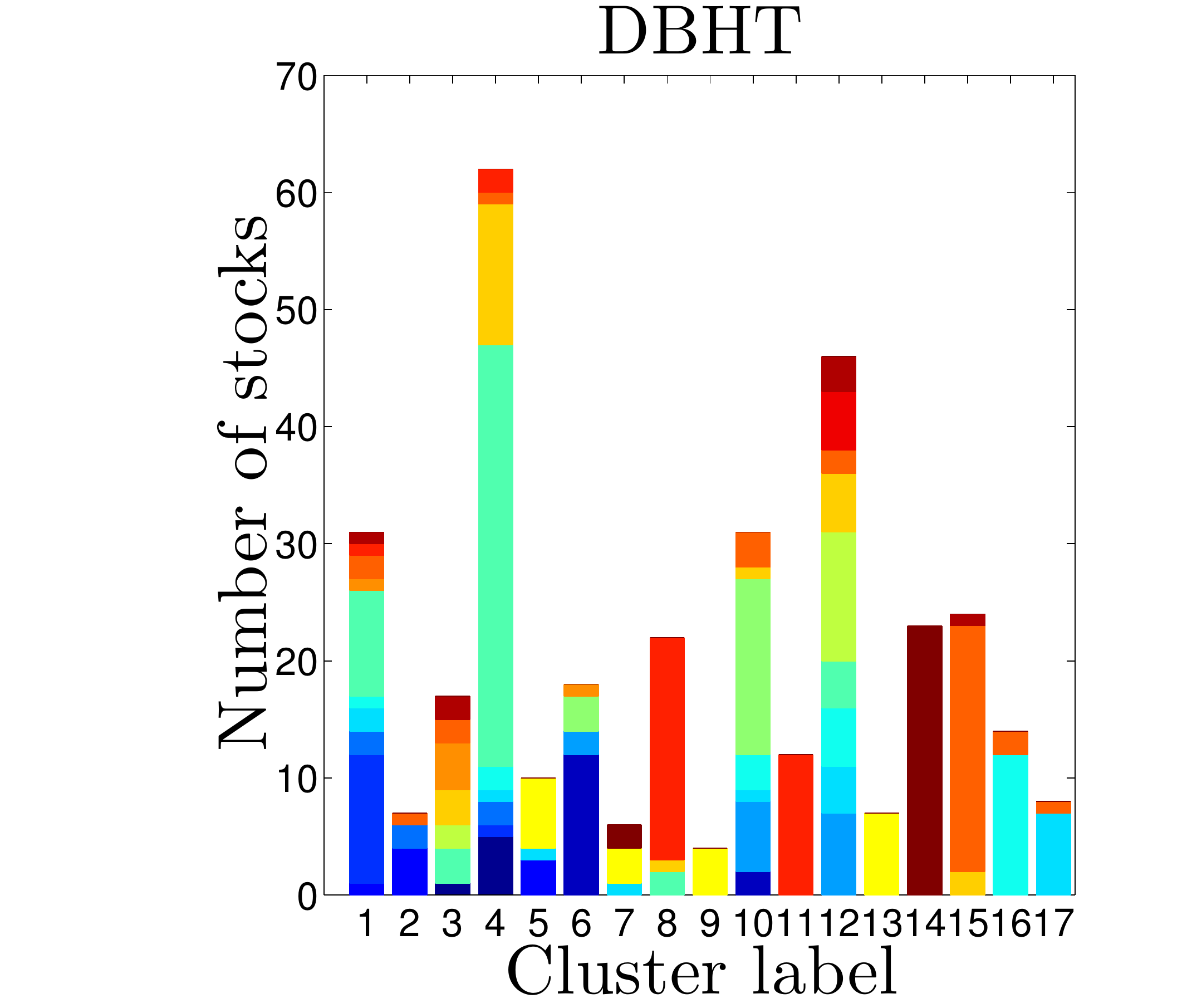}
%   \hspace{-2em}
%   \includegraphics[scale=0.4]{figure/bar_cl_composition_dbht2_subtracted.eps}
\end{tabular}
\end{center}
 \caption{\label{fig:histo_cl_composition_dbht} {\bf Number of stocks and composition of DBHT clusters in terms of ICB supersectors, on non-detrended log-returns}. The composition is shown by 
 using different colours. } 
\end{figure} 

  In Fig. \ref{fig:histo_cl_composition_dbht} we report a graphical summary of the clusters obtained applying the DBHT method to the whole time window of data (1997-2012), by using non-detrended log-returns.
  
  The DBHT returns a number of clusters, $N_{cl}$, equal to 17.
   Cluster 4, the largest, is made of 62 stocks, accounting for about the $18\%$ of the total number of stocks; cluster $9$, the smallest, contains $4$ stocks. The average size of clusters is $20.1$ stocks.
   As we can see, four clusters show a composition of stocks belonging to only one ICB supersector
  : cluster 9 and 13 (Oil \& Gas), 11 (Technology) and 14 (Utilities). Similar cases are cluster 8, made of Technology stocks for more than $86\%$, cluster 15,
  within which $91\%$ 
  of stocks are from Retail, cluster 16 ($75\%$ of stocks from Health Care) and cluster 17 ($87.5 \%$ of stocks from Food \& Beverage). Moreover there are clusters that, although showing a mixed
  composition, are composed by supersectors strictly related: the number 6 is made of Banks, Financial Services and Insurance, all supersectors that the ICB gathers in the same industry (Financial) at the superior hierarchical
  step. 
 
 There are clusters that do not show an overexpression for a particular supersector or industry: this fact points out that the clustering is after all providing an information that cannot be reduced only 
 to the industrial classification. In particular clusters $1$, $3$ and $12$ have a heterogeneous composition, covering almost all the 19 supersectors and 
 with no sector dominating the others. The cluster 4 is an intermediate case, since even though it overexpresses the Industrial Goods \& Services ($75\%$), it contains stocks belonging to 9 
 different supersectors. Interestingly the largest clusters (4, 12, 1 and 10) are all among these types of ``mixed'' clusters. 
  
\subsubsection*{Other clustering compositions}
\label{sec:linkage_clusters}

\begin{figure}[ht!]
\begin{center}
\begin{tabular}{l}
  \includegraphics[scale=0.3]{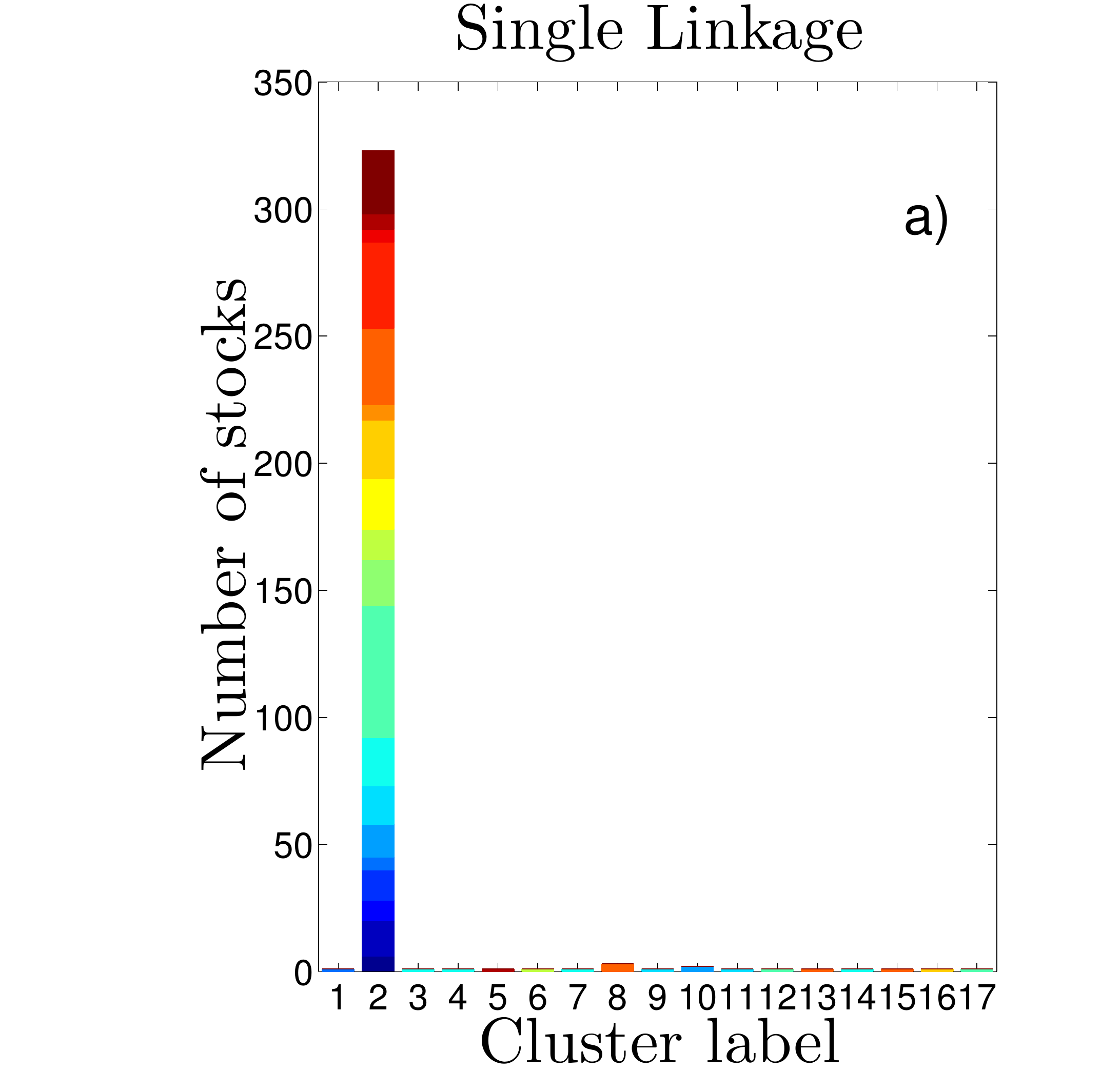}
  \includegraphics[scale=0.3]{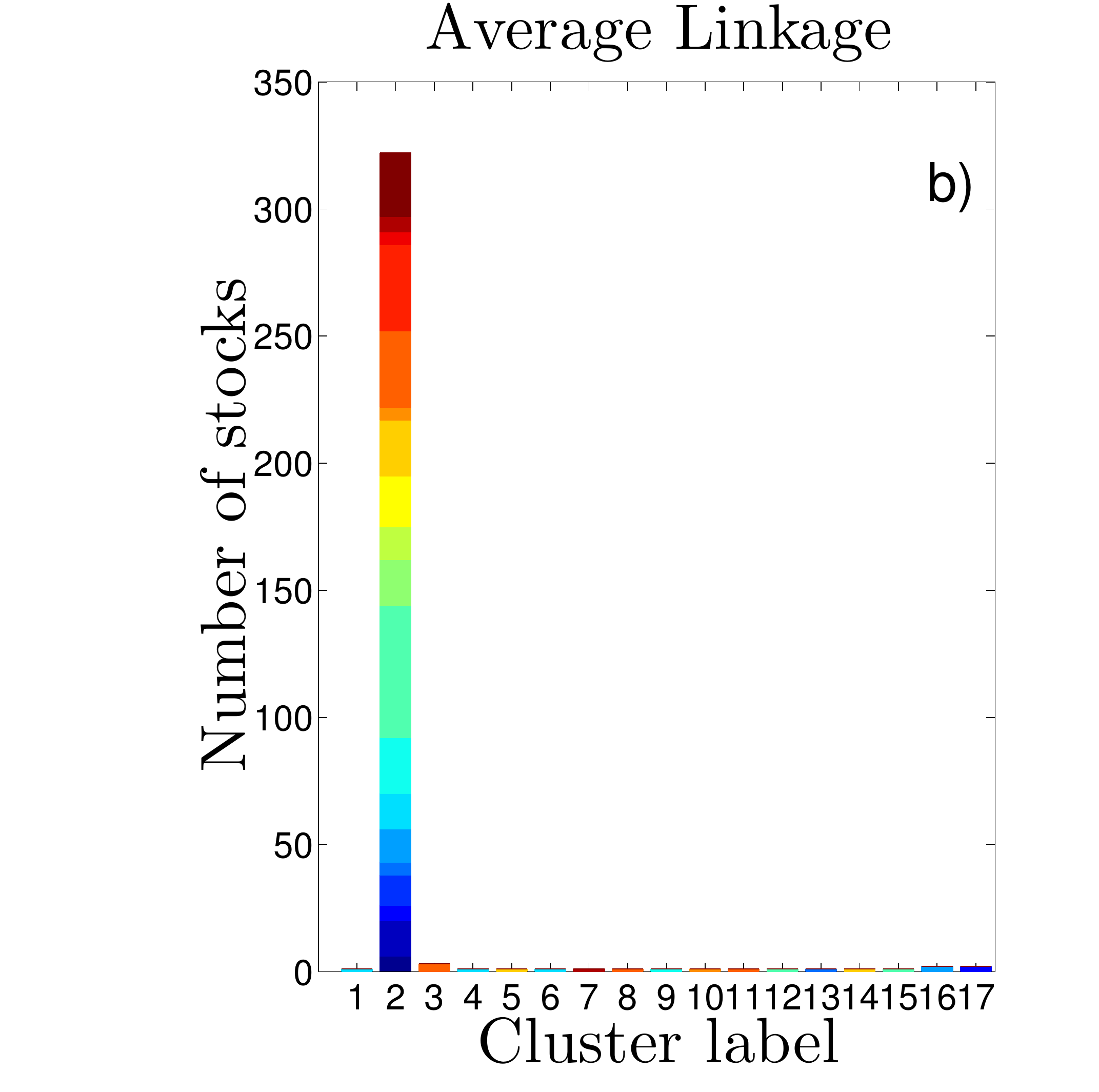}\\
  \includegraphics[scale=0.3]{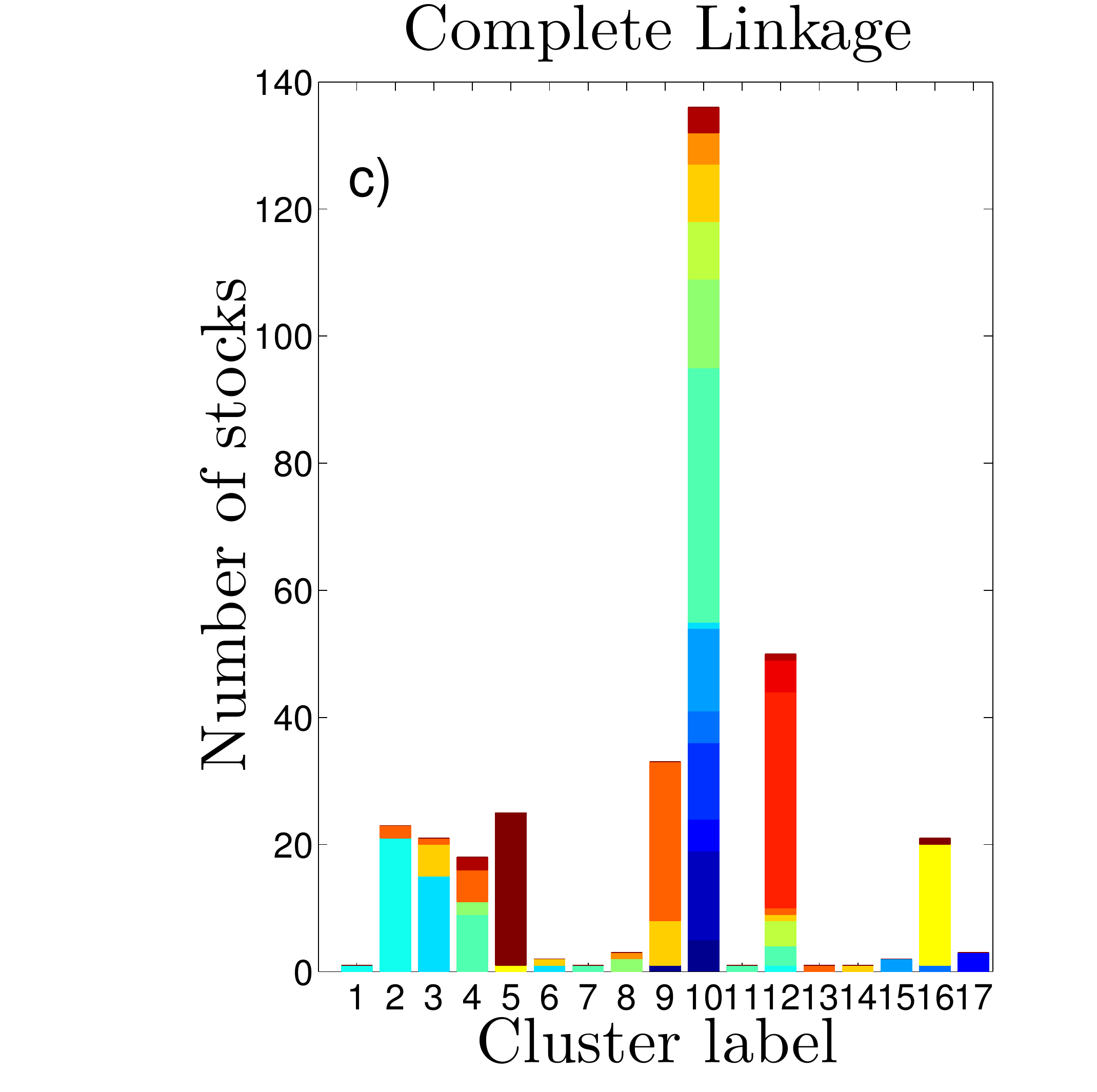}
  \includegraphics[scale=0.3]{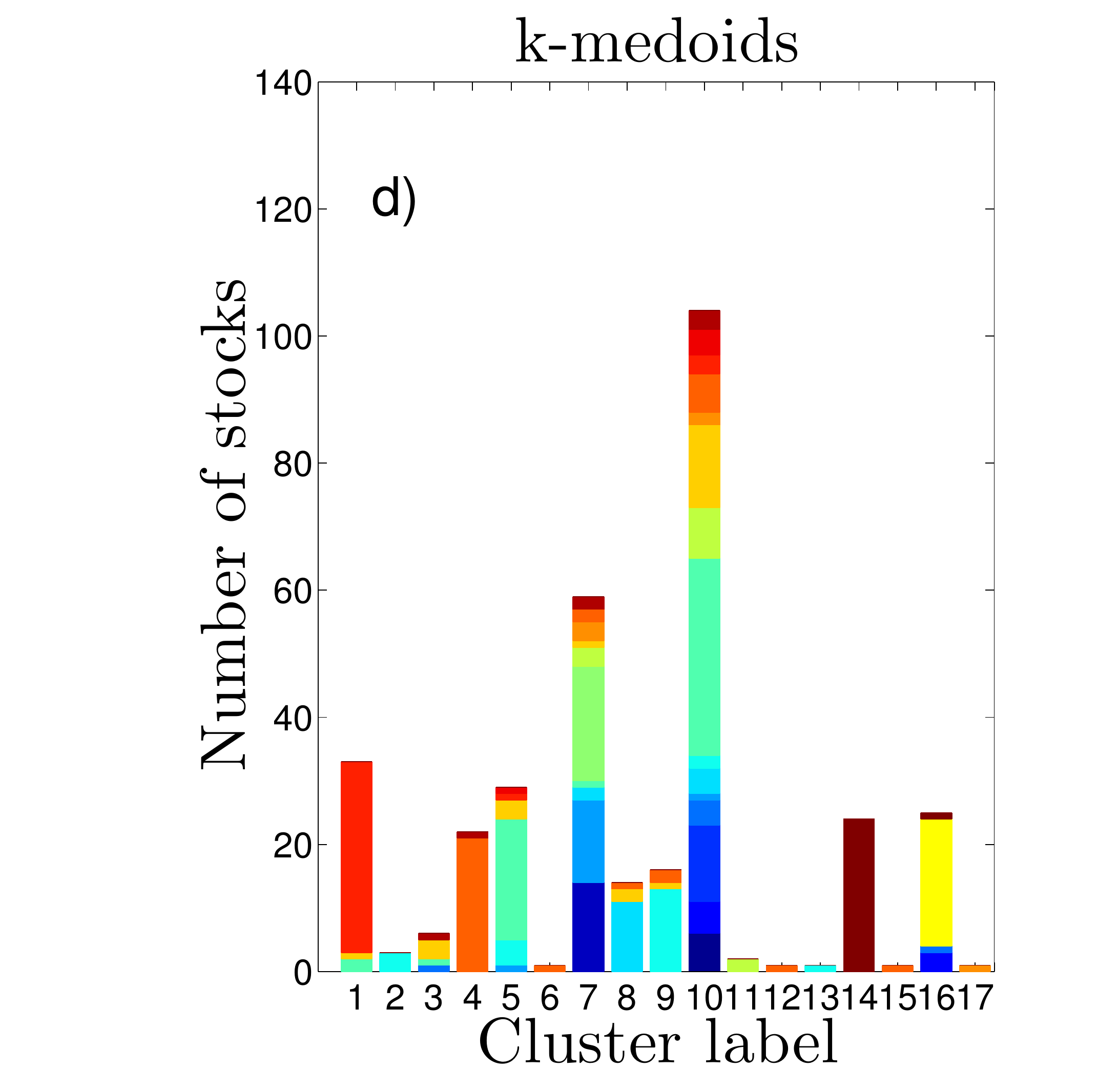}
\end{tabular}
\end{center}
 \caption{\label{fig:histo_cl_composition_linkage} {\bf Composition of clustering in terms of ICB supersectors, for different clustering methods, on non-detrended log-returns}.
 The x-axis represents the single cluster labels, the y-axis the number of stocks in each cluster. Each colour corresponds to
 an ICB supersector (the legend is the same as in Fig. \ref{fig:histo_cl_composition_dbht}. The graphs show the results for a) SL clustering, b) for AL, c) for CL and d) for k-medoids.} 
\end{figure} 

We here apply other clustering methods on the same data and compare results with DBHT clustering. The clustering methods considered are Single Linkage (SL), Average Linkage (AL), Complete Linkage (CL) and k-medoids. 
  The latter is not a
  hierarchical clustering method, so it does not provide a dendrogram: however we analysed it to compare our results with a well established clustering method.
The number of clusters, that unlike the DBHT, is a free-parameter for these methods, has been chosen equal to 17 in these cases, in order to compare the bar graphs with the Fig. \ref{fig:histo_cl_composition_dbht} for DBHT.
We plot in Fig. \ref{fig:histo_cl_composition_linkage} a), b), c) and d) the clusters compositions obtained by using these four clustering methods, namely SL, AL, CL and k-medoids.
 
 First of all we can observe that for each of them there is a strong heterogeneity in the size of clusters: SL and AL display two huge clusters of 323 and 322 stocks respectively (almost identical, having 318 stocks in common),
 with the other clusters made of one, two or three stocks. For both the algorithms this giant cluster contains stocks of all ICB sectors. 
 
 For the CL and the k-medoids the situation is quite different. For the CL, the giant cluster (cluster number 10) is much reduced in size (136 stocks), with also other three clusters (the number 12, 9 and 5)
 containing a relevant number of stocks (50, 
 33 and 25 respectively): the main supersectors that are overexpressed are Technology (cluster 12), Utilities (cluster 5), Retail (cluster 9), Oil \& Gas (cluster 16) and Health Care (cluster 2). A very similar structure occurs with
  the k-medoids, but with the giant cluster splitting further in two large clusters (7 and 10). 
 However the DBHT clustering is the one showing the largest degree of homogeneity in size and overexpression of ICB supersectors, at least for this number of clusters (see Fig. \ref{fig:histo_cl_composition_dbht} for comparison).
 
 Comparing these results with the same analyses on detrended log-returns (Fig. 3 and 4 in the paper) we can conclude that the subtraction of the market mode makes
  all the clusterings methods (with the exception of SL) more homogenous in size and more able to retrieve the ICB partition. The SL clustering instead does not seem to 
  be sensitive to this subtraction, and keeps not overexpressing any ICB supersector even in the detrended case (Fig. 4 a)). 
 
\subsection*{S4. Bootstrapping test of robustness}
\label{appendix:bootstrap}
 The basic idea of the Bootstrapping technique is the following \cite{bootstrapping}: suppose, for a given time window of length $L$, we have $N$ time series (one for each stock), each one having length $L$ .
 We can fit this data in a $N \times L$ matrix, say $X$, and calculate the correlation matrix for it, say $\rho$ ,
and a clustering using the DBHT, say $Y$. Now let us create a replica $X'$ of the matrix $X$, such that each row of $X'$ is drawn randomly among the rows of $X$, allowing multiple drawings of the same rows. From $X'$ we can again calculate 
 a correlation matrix $\rho'$ and a clustering $Y'$. 
 
 By repeating this procedure $n_{boot}$ times, we end up with $n_{boot}$ replica of clusterings, each one slightly different from the original one due to the differences between $X$ and its
replicas. This sample of replicas can be used to test the robustness of any quantity measured in the original clustering $Y$, e.g. the number of clusters. This can be done by checking whether the original measure is compatible 
with the distribution of replicas, performing e.g. a statistical hypothesis test.

 \begin{figure}[h!]
\begin{center}
\begin{tabular}{l}
  \includegraphics[scale=0.22]{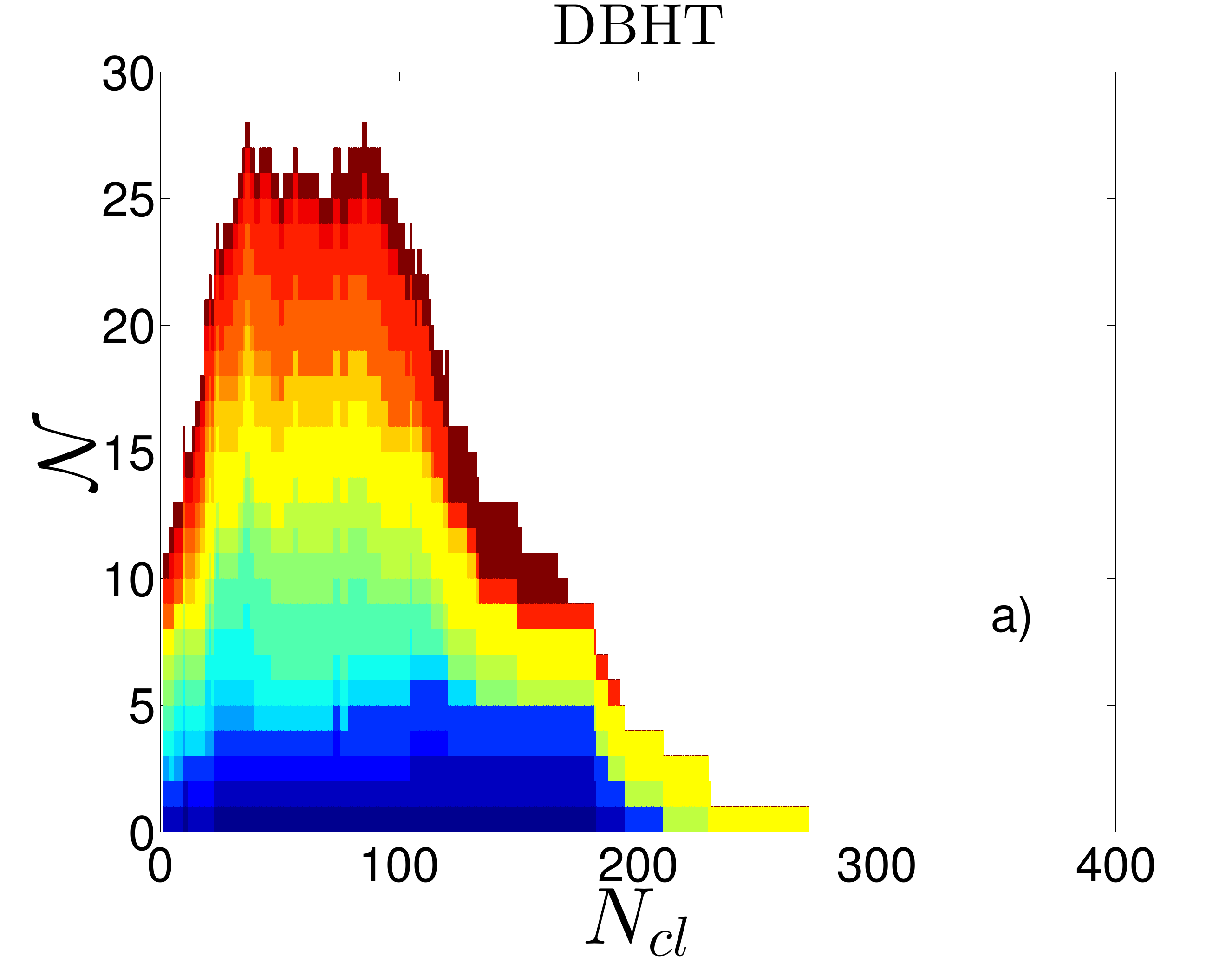}
  \includegraphics[scale=0.22]{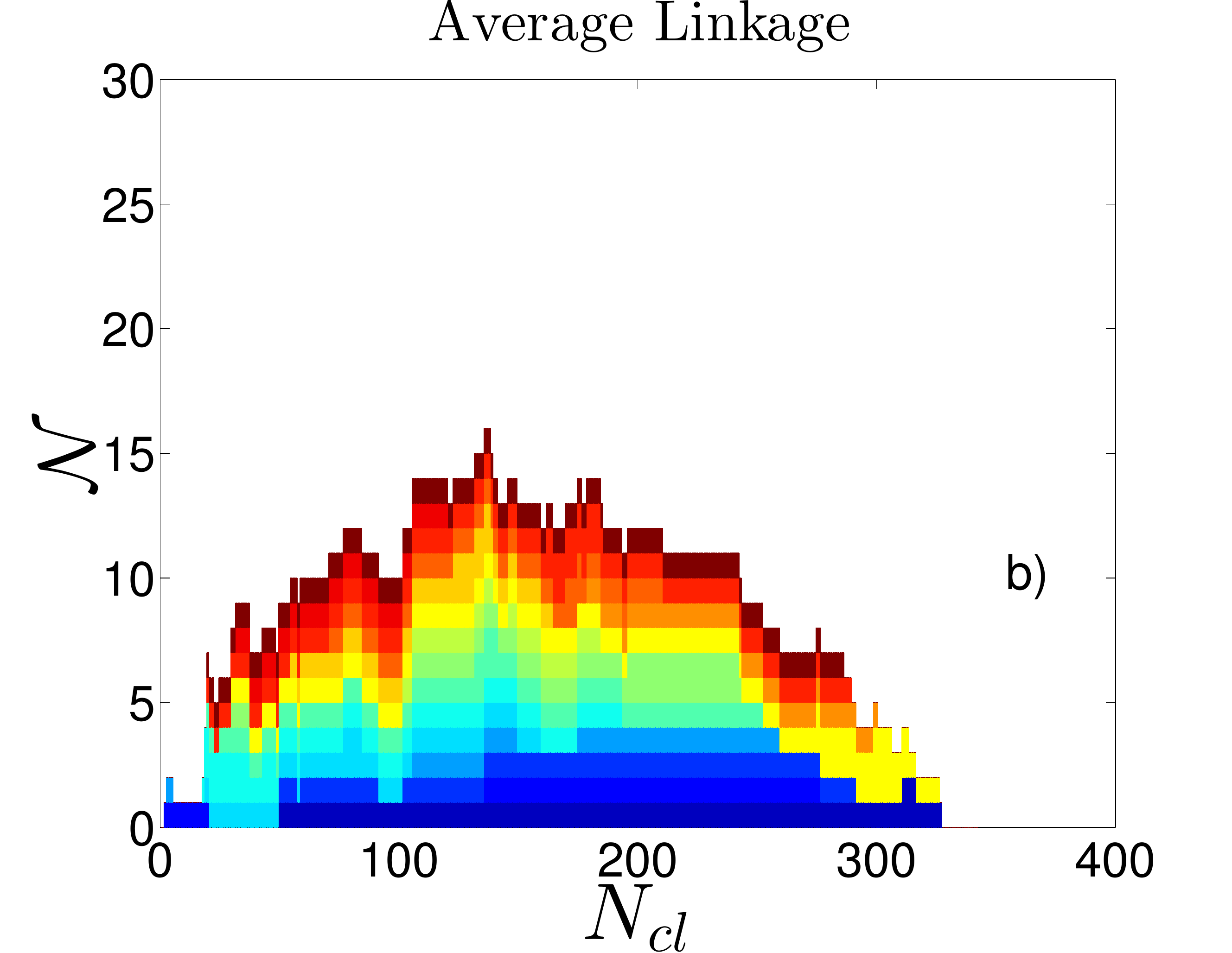}\\
  \includegraphics[scale=0.22]{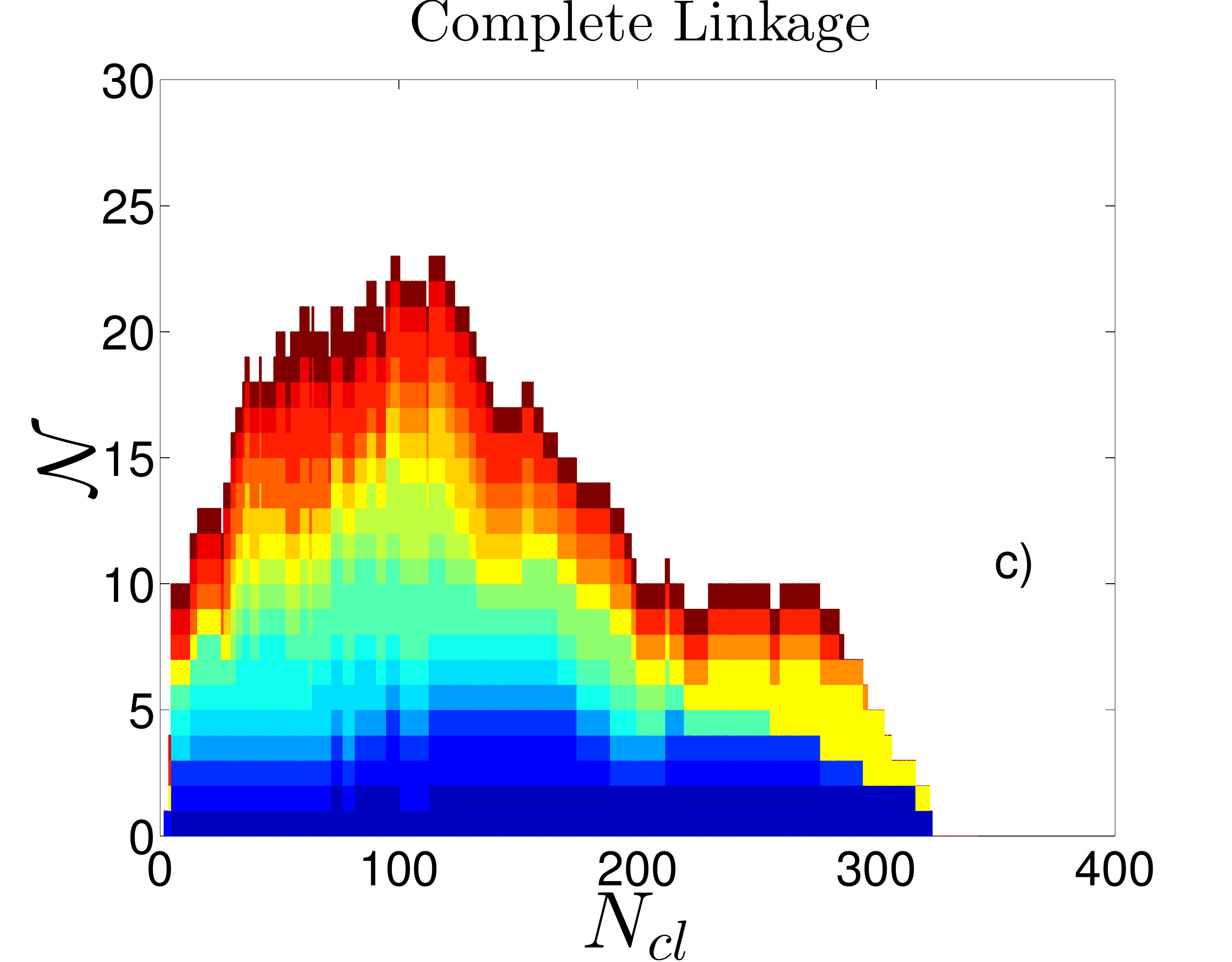}
  \includegraphics[scale=0.22]{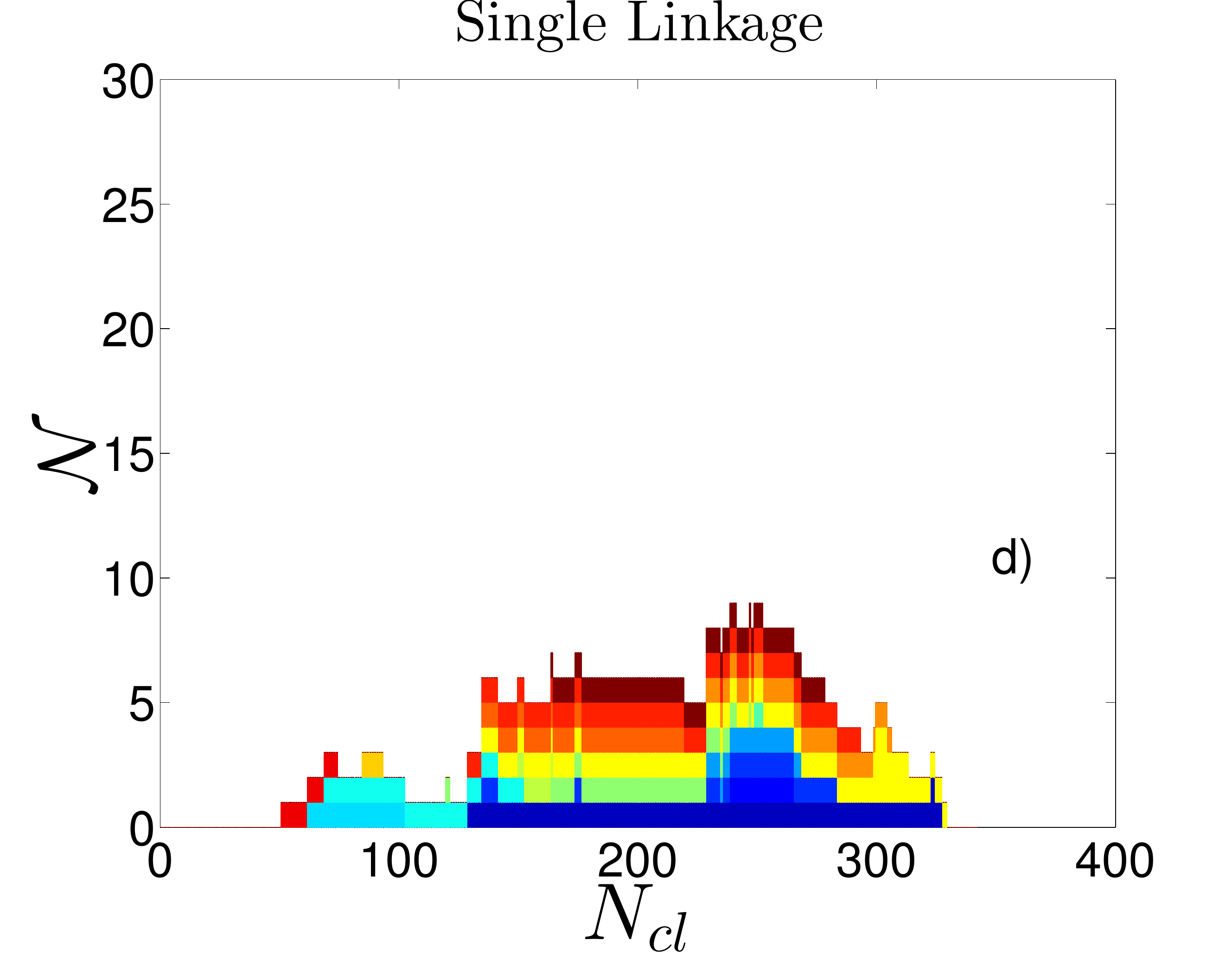}\\
  ~~~~~~~~~~~
  \includegraphics[scale=0.22]{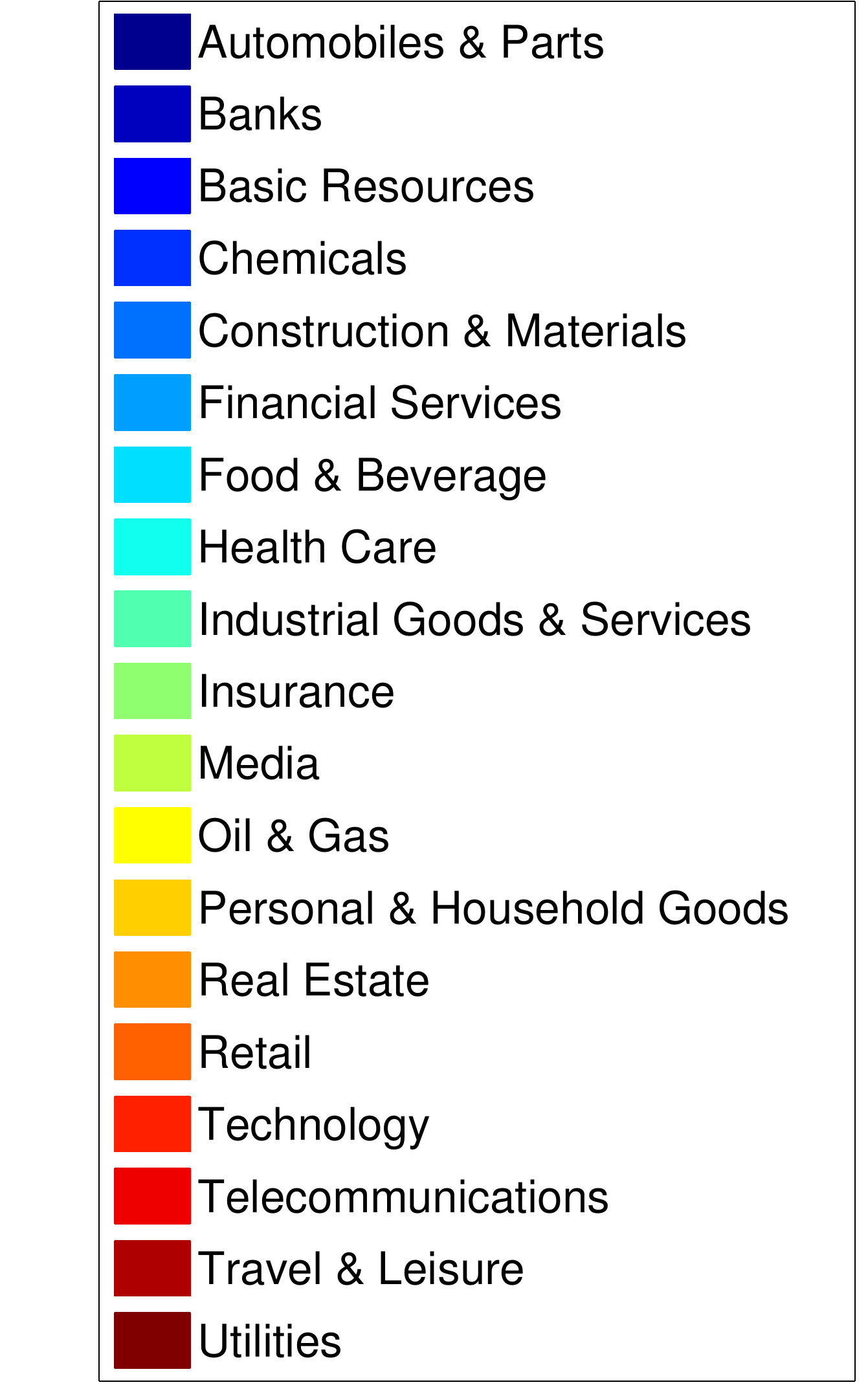}
  ~~~~~~~~~~~~~~~
  \includegraphics[scale=0.22]{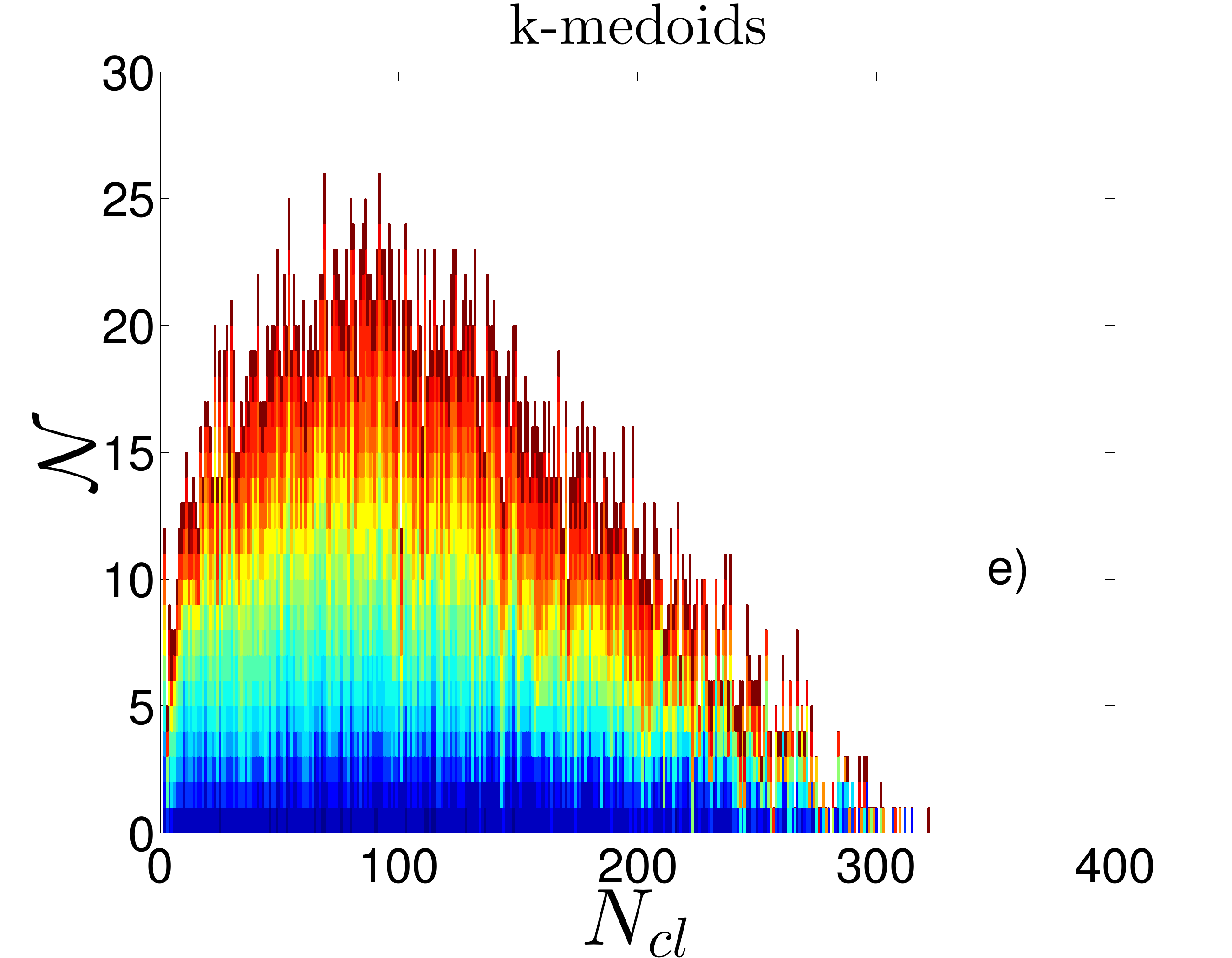}
\end{tabular}
\end{center}
 \caption{\label{fig:histo_sect_ncl_supersectors} {\bf ICB supersectors overexpression at different levels of the hierarchies}. Each bar graph shows, varying the number of clusters $N_{cl}$, how many times ($\mathcal{N}$) 
 an ICB supersector is overexpressed by a cluster, according to the Hypergeometric hypothesis test
 (i.e., number of tests being rejected). Each colour shows the number of overexpressions for each ICB supersector. In graphs a)-e) the results for DBHT, AL, CL, SL and k-medoids clustering respectively are shown.  
 The correlations are calculated on non-detrended log-returns.} 
\end{figure} 

 \begin{figure}[h!]
\begin{center}
\begin{tabular}{l}
  \includegraphics[scale=0.3]{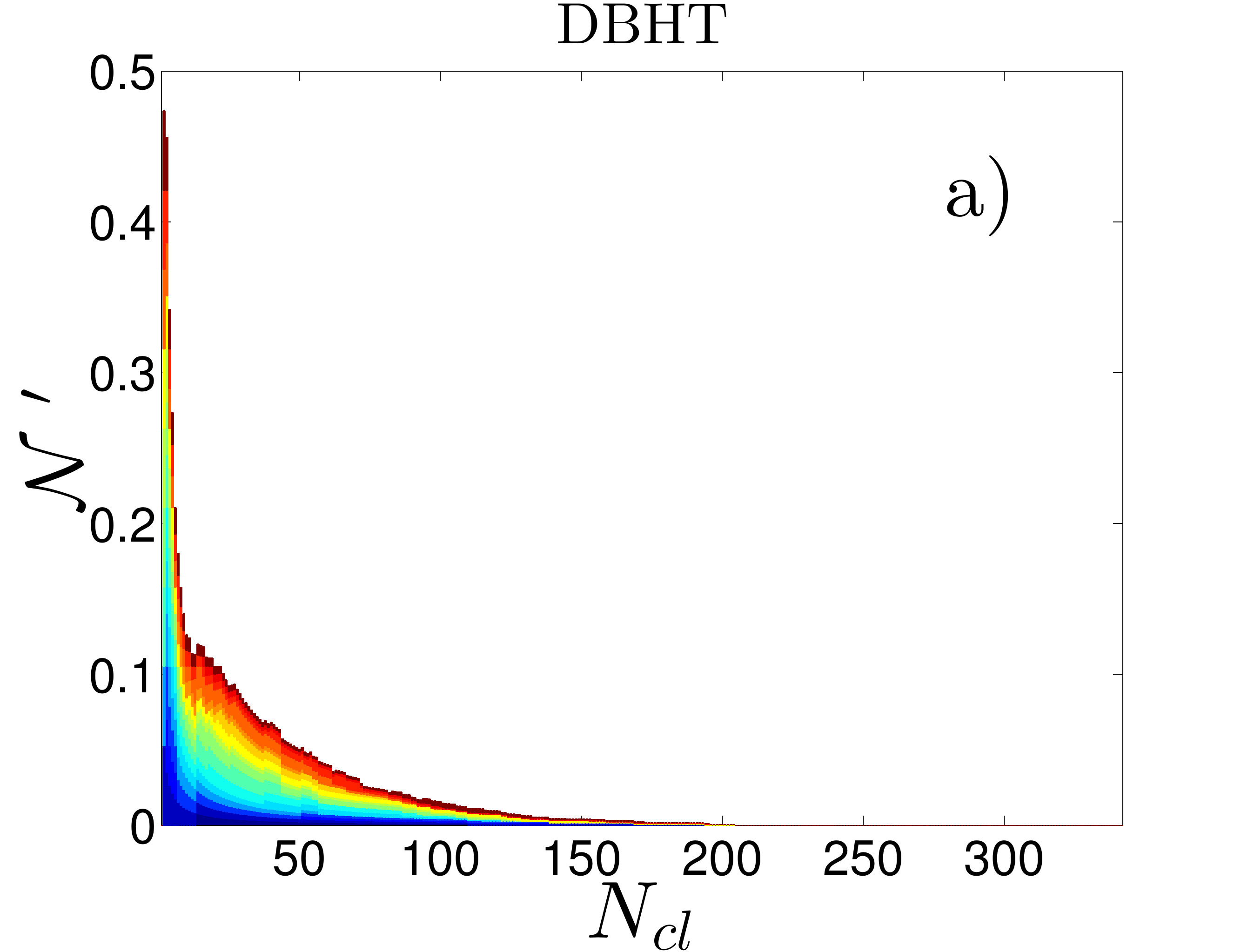}
  \includegraphics[scale=0.3]{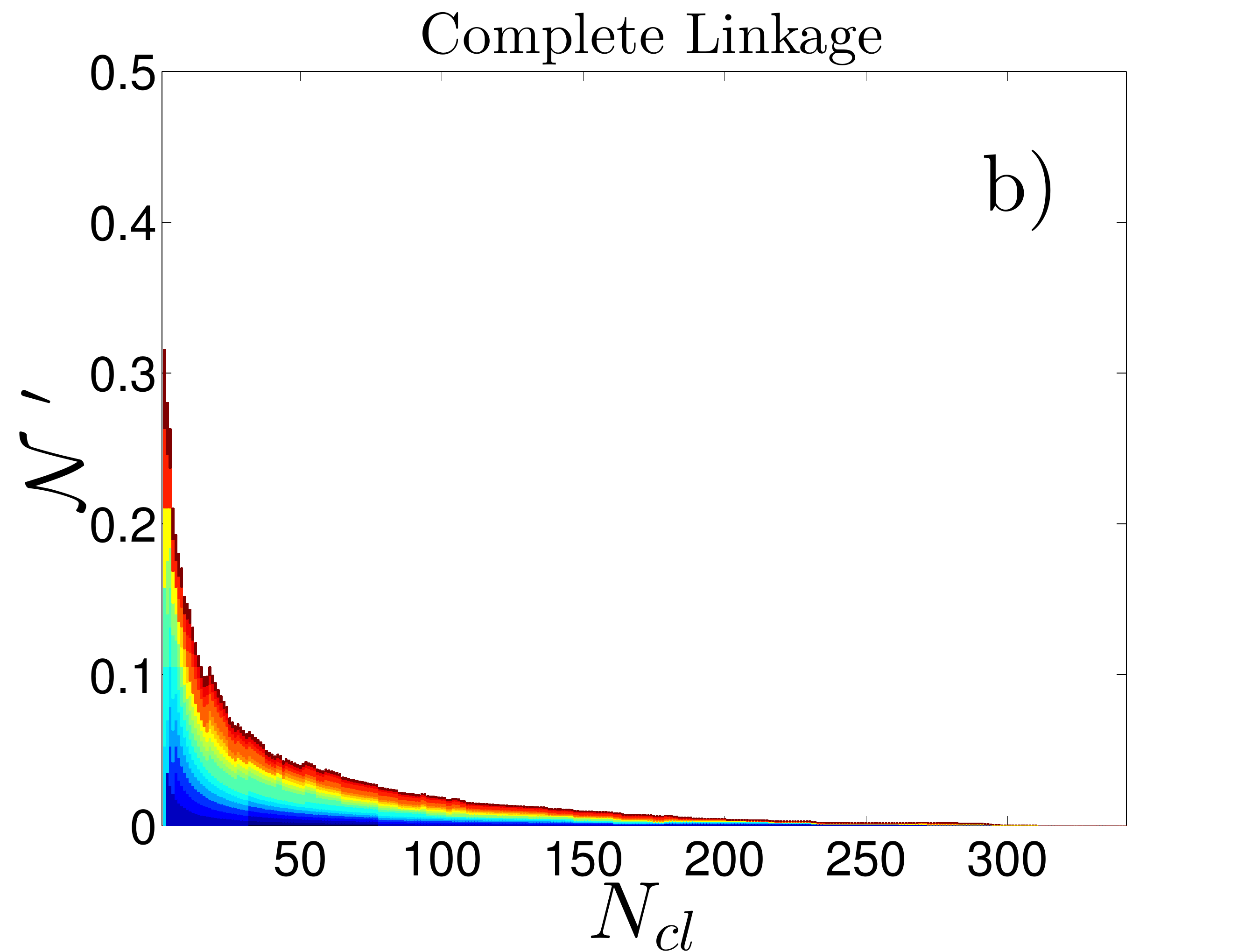}
\end{tabular}
\end{center}
 \caption{\label{fig:histo_sect_ncl_supersectors} {\bf ICB supersectors overexpression as percentage of pairs cluster/supersector rejecting the hypergeometric test}. Each bar graph shows,
 varying the number of clusters $N_{cl}$, how many times ($\mathcal{N}$) 
 an ICB supersector is overexpressed by a cluster according to the Hypergeometric hypothesis test
 (i.e., number of tests being rejected), divided by the total number of Hypergeometric tests performed ($0.5 \times N_{cl} \times N_{ICB}$, with $N_{ICB}$ the number of ICB supersectors): $\mathcal{N'} = \frac{2\mathcal{N}}{N_{cl} \times N_{ICB}}$.
 Each colour shows the number of overexpressions for each ICB supersector. In graphs a)-b) the results for DBHT and CL clusterings respectively are shown.  
 The correlations are calculated on detrended log-returns.} 
\end{figure}

\end{document}